\documentclass[5p,twocolumn]{elsarticle}
\usepackage{graphics,graphicx}
\usepackage{mathrsfs}
\usepackage{amsmath,amsfonts,amssymb,wasysym}
\usepackage{color}
\usepackage{natbib}
\usepackage{epstopdf}
\usepackage{epsfig}
\usepackage{bm}
\usepackage{listings}
\usepackage{graphicx}
\usepackage{newtxtext}
\usepackage{newtxmath}
\usepackage{hyperref}
\usepackage{subcaption}
\usepackage{comment}
\usepackage{cuted}

\biboptions{longnamesfirst,sort&compress}

\hypersetup{
    colorlinks = true,
    urlcolor   = blue,
    citecolor  = black,
}

\newcommand{\RomanNumeralCaps}[1]
\linenumbers

\def\half{\mbox{$1\over2$}}

\DeclareMathOperator{\sgn}{\mathrm{sgn}}

\DeclareMathOperator{\mW}{\mathcal{W}}
\DeclareMathOperator{\mR}{\mathcal{R}}
\DeclareMathOperator{\mG}{\mathcal{G}}

\DeclareMathOperator{\mS}{\mathcal{S}}

\DeclareMathOperator{\mJ}{\mathcal{J}}
\DeclareMathOperator{\mf}{\epsilon}

\journal{Journal of Non-Newtonian Fluid Mechanics}

\begin{document}

\begin{frontmatter}

\title{Air-driven dynamics of viscoplastic liquid layers}

\author[ad1]{J. D. Shemilt\corref{cor1}} 
\author[ad1]{N. J. Balmforth} 
\author[ad2]{D. R. Hewitt}

\address[ad1]{ Department of Mathematics, University of British Columbia,
  Vancouver, BC, V6T 1Z2, Canada}
    \address[ad2]{Department of Applied Mathematics and Theoretical Physics,
University of Cambridge, Wilberforce Road, Cambridge CB3 0WA, UK}
\cortext[cor1]{Corresponding author: {\it E-mail:} shemilt{@}math.ubc.ca}

\date{\today}

\begin{abstract}
Airway clearance by coughing is a key mechanism for mucus transport, particularly in obstructive lung diseases associated with altered mucus rheology. We investigate the dynamics of a viscoplastic liquid film driven by flow in a turbulent air layer, which is a model for air-driven mucus transport that incorporates yield-stress effects. Our theoretical analysis is based on a long-wave model for the liquid film flow, and we complement this with experiments, in which layers of Newtonian and yield-stress liquids are exposed to air flow in a rectangular duct. We demonstrate how perturbations to the layer depth can lead to localised yielding and wave generation. Rapid wave growth occurs when the fluid ahead of the oncoming wave is unyielded, so that as the wave propagates, it consumes this static fluid while depositing a much thinner film behind. This mechanism causes dramatic ``blow-out'' events in experiments, where liquid hits the roof of the tank. By contrast, in Newtonian thin films, multiple surface waves typically form, and blow-out only occurs in experiments when a Newtonian film is sufficiently thick. 

\end{abstract}


\end{frontmatter}

\numberwithin{equation}{section}

\section{Introduction}

Coughing is a mechanism for clearance in which the mucus layer lining an
airway is exposed to high-velocity air flow. This mechanism can become
impeded in lung diseases such as cystic fibrosis and chronic obstructive
pulmonary disease, in which mucus yield stress can be significantly
increased over that of healthy mucus \cite{patarin_rheological_2020},
reducing the effectiveness of air-driven transport.
Treatments of those diseases include
airway clearance techniques that incorporate forced expirations,
aiming to induce airway mucus transport via air flow \cite{belli2021airway}. 

As a model for air-flow-driven mucus transport,
King et al. \cite{king1985clearance} conducted an experimental study
in which they exposed a layer of gel composed of locust bean gum to high-velocity air flow in a rectangular duct.
They focused on quantifying the efficiency of liquid transport (as measured by motion of tracer particles), finding reduced rates of transport when the liquid layer was thinner or when there was a higher concentration of gum in the gel. Further experimental studies followed using their `simulated-cough machine', in which liquids with viscoelastic or thixotropic rheologies were used \cite{king1987role,zahm1991role}, the gel layer was lubricated with a lower viscosity sub-layer \cite{zahm1989role}, or repeated bursts of air flow were applied \cite{zahm1991role}. More recently, there have also been experiments conducted in `simulated-cough machines' focusing on how droplets may be generated by bursts of air flow over a liquid film \cite{kant2023bag,li2025viscoelasticity}.

Basser et al. \cite{basser_1989_cough}, using a similar experimental setup but with mayonnaise as their working liquid, focused more closely on the mechanism by which waves are generated on a layer of non-Newtonian fluid by air flow in a confined duct. They identified that, when the air flow was strong enough to initiate motion in the mayonnaise, an isolated surface wave would typically form, then rapidly grow as it was transported forwards, until liquid made contact with the roof of the duct and a dramatic blow-out of liquid from the tank occurred. The dynamics were qualitatively different than what they observed when a Newtonian liquid was used: in that case, multiple surface waves formed, which propagated more slowly and which were less likely to trigger blow-out.
Although these authors suggested that a yield stress was important for
explaining wave growth in mayonnaise, the detailed mechanism they
hypothesized for wave formation was a little different:
they argued that an avalanche-like instability arose in which liquid would
suddenly begin slipping along the base of the duct, leading to thickening
of the layer and downstream transport. Basser et al. provided evidence
that some wall slip featured in their experiments by covering the base of their tank with a rougher material. Whilst this alteration did impact the critical air-flow rate required to trigger motion, it did not qualitatively affect the wave dynamics. Their imaging of the air-liquid interface was also limited, and their theoretical analysis identified only the yielding threshold for a uniformly thick layer. It is therefore not clear that either avalanche-like failure or slip are key in prompting dramatic wave growth on yield-stress liquid layers.

A number of previous studies have focused on two-layer flows
with Newtonian fluids. On the experimental side, Jurman \& McCready \cite{jurman_study_1989} examined thin viscous films sheared by air flow, finding regular
periodic surface waves forming at lower air flow rates and solitary waves
emerging as the flow rate was increased.
The formation of solitary waves on relatively deep viscous liquid layers has also been studied experimentally \cite{paquier2016viscosity,aulnette2019wind,aulnette2022kelvin}.

On the theoretical side,
Matar et al. \cite{matar2007interfacial} derived a 
model based on a long-wave approximation
of the equations for two-layer Newtonian fluid flow, assuming the lower layer to be inertialess and the flow in the upper layer to be quasi-steady and laminar. They computed travelling-wave solutions to their evolution equation,
and studied wave dynamics in long domains where multiple waves could
form and complex wave interactions could occur. Solutions 
exhibited finite-time blow-up, where the lower fluid
layer rapidly approached the top boundary, when the mean layer
thickness was sufficiently high or when multiple waves coalesced.
Similar models have been proposed and explored
for wave formation
on liquid layers on inclines below deep potential flow
\cite{king1993air,meng2024steady,meng2025dynamics},
the viscous lining inside a cylindrical tube
\cite{camassa2012ring,camassa_2017_air}, or liquid films coating the underside of a horizontal plane \cite{hu2024nonlinear}.


Models with a single evolution equation for the very viscous lower layer,
like those used by Matar et al. \cite{matar2007interfacial} or
Meng et al. \cite{meng2024steady,meng2025dynamics}, benefit from
their relative simplicity. However, models of this type may be open to some criticism, particularly in their propensity for finite-time blow-up, and given that the treatment of the flow in the less viscous upper fluid
can be relatively simplistic. Other studies of gas-liquid flows
aimed at treating the flow in both layers in more detail either simulate the full two-dimensional laminar problem \cite{frank2008numerical}
or lead to coupled
evolution equations for the liquid layer flow 
\cite{tseluiko2011nonlinear,dietze2013wavy,ishimura2023gas,hu2025surface}.
Although the simpler models 
may fail to incorporate all the physics of air-liquid interaction,
they can still predict surface-wave instability.
As such, this type of model offers a convenient avenue to explore
the surface-wave instability of a layer of yield-stress fluid
driven by air flow, for which the material's rheology
may significantly complicate the theoretical description.

Moriarty \& Grotberg \cite{moriarty1999flow}, motivated by modelling airway mucus clearance, examined the
linear stability of a viscoelastic slab driven by air flow over
a viscous sub-layer. There have also been several studies focusing on CFD models of cough, which have assumed Newtonian \cite{paz2017cfd,paz2019analysis,ren2020numerical} or shear-thinning generalised Newtonian \cite{yi2021computational} rheological models for the mucus layer. These CFD studies largely focused on quantifying relatively simple measures of liquid transport efficiency as functions of layer depth and viscosity, rather than providing detailed examinations of the dynamics in the liquid layer. Kant et al. \cite{kant2023bag} also conducted CFD simulations of a Newtonian air-driven flow, but focusing on the process of aerosol generation via the formation of thin bag-like structures from the liquid layer. This work was recently extended to examine the effects of viscoelasticity on the aerosol generation process \cite{li2025viscoelasticity}, with experiments again being complemented by numerical simulations of droplet fragmentation. 


Theoretical models of some other mucus flows have incorporated non-Newtonian
rheological effects, including viscoplasticity. 
Shemilt et al. \cite{shemilt_2022_surface,shemilt2023surfactant,shemilt2025viscoplasticity} demonstrated how viscoplasticity can stabilize the surface-tension-driven instability of a film coating a cylindrical tube in the absence of air flow.
Erken et al. \cite{erken_2022_elastoviscoplastic}
and Fazla et al. \cite{fazla2024effects}
conducted full two-dimensional simulations for the same surface-tension-driven flow, but using more complicated rheological models for the liquid.

Our focus in the present study is to investigate the detailed dynamics of wave generation in a yield-stress liquid driven by overlying air flow. We use long-wave modelling, complemented by experiments with yield-stress and Newtonian fluids in which the evolution of the liquid interface is captured in more detail than was possible in \cite{basser_1989_cough}. Whilst the primary motivation for the study arises from modelling airway mucus transport, there may be additional applications in engineering or geophysical processes where a low-viscosity fluid flows over a non-Newtonian fluid film \cite{landel21,charru2013sand}.
The structure of shallow viscoplastic flows has been outlined in a
number of previous studies (see the review \cite{balmforth_yielding_2014}),
including several studying free-surface flows with surface tension
\cite{jalaal_long_2016,jalaal_stoeber_balmforth_2021,van_der_kolk_tieman_jalaal_2023,shemilt_2022_surface,shemilt2023surfactant,shemilt2025viscoplasticity}.


The paper is outlined as follows. 
We present the long-wave model in \S\ref{sec:model},
and interrogate the dynamics it captures in \S\ref{sec:res_model}.
In \S\ref{sec:expts}, we describe our experiments, in which layers of Newtonian or yield-stress liquids are sheared by air flow inside a rectangular duct.
Finally, we discuss our findings, drawing comparisons between theoretical and experimental results, in \S\ref{sec:discussion}.

\section{Long-wave model formulation}\label{sec:model}

\begin{figure}
    \centering    \includegraphics[width=.9\linewidth]{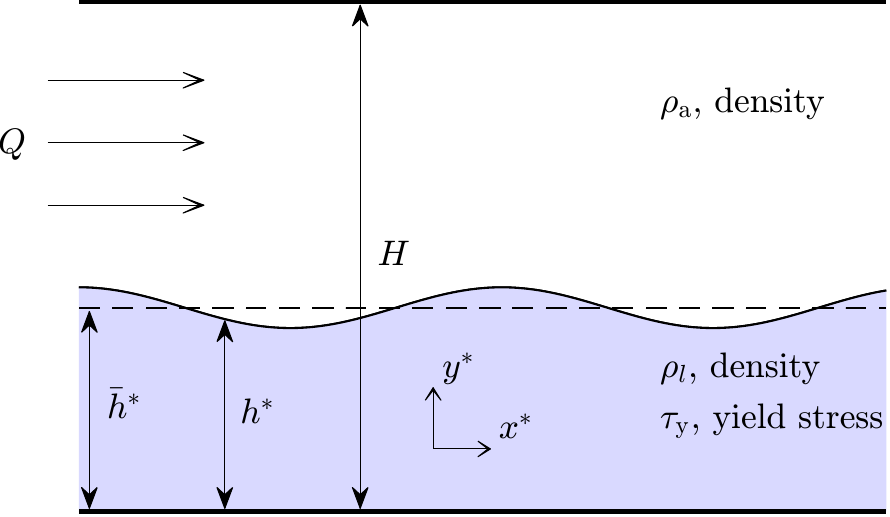}
    \caption{Sketch of the long-wave model geometry.}
    \label{fig:geomsketch}
\end{figure}

As sketched in figure \ref{fig:geomsketch},
we consider a two-dimensional channel of height $H$
containing a layer of
liquid above which a gap permits air flow. We use a Cartesian coordinate system
$(x^*,y^*)$ to describe the geometry; the channel is bounded
by rigid walls at $y^*=0$ and $y^*=H$, with the liquid layer
occupying $0<y^*\leq h^*(x^*,t^*)$, and air filling the gap $h^*<y^*<H$.

\subsection{Air flow}\label{sec:model_air}

We adopt a quasi-steady,
depth-average model for the air flow, in which the mean velocity
along the channel is $U^*(x^*,t^*)$ and the local pressure is $P^*(x^*,t^*)$.
Assuming that the air remains approximately incompressible, 
conservation of mass and momentum imply that
\begin{align}
    Q &= U^*(H-h^*) ,\label{SVmass}\\
    \rho_aU^*\frac{\partial U^*}{\partial x^*} &=
    - \frac{\partial P^*}{\partial x^*}
    - \frac{\tau_\mathrm{w,H}^*}{H-h^*} - \frac{\tau_\mathrm{w,h}^*}{H-h^*} ,
    \label{SVmom}
\end{align}
where $Q$ is the (constant) flux of air into channel, $\rho_a$ is the air
density, and $\tau_\mathrm{w,H}^*$  and $\tau_\mathrm{w,h}^*$ denote
the drag exerted by the upper wall, $y^*=H$, and interface, $y^*=h^*(x^*,t^*)$,
respectively.
For the latter, we adopt the turbulent Ch{\'e}zy drag law,
\begin{equation}
  \tau_\mathrm{w,H}^* = 
  \tau_\mathrm{w,h}^* =
  \tau_\mathrm{w}^* = \mf\rho_a {U^*}^2,\label{tauwDIM}
\end{equation}
with equal friction factors $\mf$
\citep[e.g.,][]{fowler2011mathematical}.
Equations \eqref{SVmass} and \eqref{SVmom} are
then analogous to the quasi-steady
Saint-Venant equations for open-channel flows \citep{fowler2011mathematical}. 
In pipe and channel flows, the friction factor $\mf$ is relatively small
and depends weakly on the Reynolds number \citep{fowler2011mathematical}
and the smoothness of the walls \cite{schlichting2016boundary}.
Here, for simplcity, we take the friction factor to be constant
and exploit $\mf\ll1$ as a small parameter in developing a
long-wave model.
We further assume that $q_0/Q\ll1$, where $q_0$ is a characteristic horizontal
flux in the liquid layer, underscoring how changes in the liquid layer height,
$h^*$, take place relatively slowly and the air flow is quasi-steady.

To non-dimensionalise \eqref{SVmass}-\eqref{tauwDIM},
we define
\begin{equation*}
    U^*= \frac{Q}{H}U,\quad h^*= Hh,\quad x^*=\frac{H}{\mf}x,
\end{equation*}
\begin{equation}
    P^*=\frac{\rho_\mathrm{a}Q^2}{H^2}P,\quad \tau^*_{\mathrm{w}}= \frac{\mf\rho_{\mathrm{a}}Q^2}{H^2}\tau_{\mathrm{w}}.
    \label{model:air_nondim}
\end{equation}
The choices in \eqref{model:air_nondim} reflect a long-wave scaling in which
horizontal length scales are much larger than vertical ones,
and a balance between turbulent drag and air inertia.
The scaled versions of \eqref{SVmass}-\eqref{tauwDIM} can then be rewritten
in the form,
\begin{eqnarray}
    - P_x &\approx& \frac{h_x + 2}{(1-h)^3} ,\label{pxAIR}\\
    \tau_{\mathrm{w}} &\approx& \frac{1}{(1-h)^2},\label{tauwAIR}
\end{eqnarray}
where we have used subscripts in $x$ as a convenient shorthand
for partial derivatives.

\subsection{Flow in the liquid layer}\label{sec:model_liquid}

Conservation of mass and momentum in the liquid layer demand
\begin{equation}
    \frac{\partial u^*}{\partial{x^*}} + \frac{\partial{v^*}}{\partial{y^*}} = 0,\label{noslipmassconsdim}
\end{equation}
\begin{equation}
    \rho_l\left(\frac{\partial{u^*}}{\partial{t^*}}+{u^*}\frac{\partial{u^*}}{\partial{x^*}} + {v^*}\frac{\partial{u^*}}{\partial{y^*}}\right) = -\frac{\partial{p^*}}{\partial{x^*}} + \frac{\partial{\tau}_{xx}^*}{\partial{x^*}} + \frac{\partial{\tau}_{xy}^*}{\partial{y^*}},\label{nosliphorizmomdim}
\end{equation}
\begin{equation}
    \rho_l\left(\frac{\partial{v^*}}{\partial{t^*}}+{u^*}\frac{\partial{v^*}}{\partial{x^*}} + {v^*}\frac{\partial{v^*}}{\partial{y^*}}\right) = -\frac{\partial{p^*}}{\partial{y^*}} + \frac{\partial{\tau}_{xy}^*}{\partial{x^*}} + \frac{\partial{\tau}_{yy}^*}{\partial{y^*}} - \rho_lg,\label{noslipvertmomdim}
\end{equation}
where $\rho_l$ is the liquid density,
the velocity is $(u^*,v^*)$, the pressure is $p^*$,
gravitational acceleration is $g$, and $\boldsymbol{\tau}^*$ denotes the deviatoric stress tensor.

At the free surface, the kinematic condition is
\begin{equation}
    \frac{\partial{h^*}}{\partial{t^*}} + {u^*}\frac{\partial{h^*}}{\partial{x^*}} = {v^*}\quad\mbox{at}\quad {y^*} = {h^*},\label{noslipkinBC}
\end{equation}
and stress continuity implies that
\begin{equation}
    \boldsymbol{\hat{n}}\cdot(-{p^*} + \boldsymbol{\tau}^*)\cdot\boldsymbol{\hat{n}} = -{P^*} + \sigma{\kappa^*}\quad\mbox{at}\quad {y^*} = {h^*},\label{noslipnormstressBCdim}
\end{equation}
\begin{equation}
    \boldsymbol{\hat{t}}\cdot(-p^* + \boldsymbol{\tau}^*)\cdot\boldsymbol{\hat{n}} = {\tau}_\mathrm{w}^*\quad\mbox{at}\quad {y^*} = {h^*},\label{noslipshearstressBCdim}
\end{equation}
where $\boldsymbol{\hat{n}}$ and $\boldsymbol{\hat{t}}$ are unit normal and tangent vectors to the free surface, respectively, ${\kappa}^*$ is the free-surface curvature and $\sigma$ is the surface tension, assumed to be constant. At the base of the layer, we assume no slip,
\begin{equation}
    {u^*} = {v^*} = 0 \quad\mbox{at}\quad {y^*} = 0.\label{noslipbaseBCs}
\end{equation}

We take 
the liquid to be a viscoplastic fluid obeying the Bingham
constitutive law,
\begin{equation}
\begin{array}{ll}
   {\tau}_{ij}^* = 
       \left(\eta + \frac{\tau_\mathrm{y}}{{\dot\gamma}^*}\right){\dot\gamma}_{ij}^*  & \quad\mbox{if}\quad{\tau^*} \geq \tau_\mathrm{y} \\
       {\dot{\gamma}}_{ij}^* = 0 & \quad \mbox{if}\quad{\tau^*} < \tau_\mathrm{y},
\end{array}
\label{stokesconstit}
\end{equation}
where $\tau_\mathrm{y}$ is the yield stress, $\eta$ is the plastic viscosity, 
\begin{equation}
  \tilde{\dot{\gamma}}_{ij} = \frac{\partial u_i^*}{\partial x_j^*}
 +   \frac{\partial u_j^*}{\partial x_i^*},\label{gammadotdefn}
\end{equation}
and $\tilde{\tau}$ and $\tilde{\dot\gamma}$ denote second tensor invariants.
%

Further to \eqref{model:air_nondim}, we non-dimensionalise \eqref{noslipmassconsdim}-\eqref{gammadotdefn} by defining
\begin{equation*}
    y^* = yH,\quad t^* = \frac{\eta H^2t}{\rho_\mathrm{a}\mf^2Q^2},\quad (u^*,v^*) = \frac{\rho_\mathrm{a}\mf Q^2}{\eta H}\left(u,\mf v\right),
    \end{equation*}
    \begin{equation} 
    p^* = \frac{\rho_\mathrm{a}Q^2{p}}{H^2 }, \quad 
    \left({\tau}_{xx}^*,{\tau}_{xy}^*,{\tau}_{yy}^*\right) = \frac{\rho_\mathrm{a}\mf Q^2}{H^2}\boldsymbol{\tau}, \quad
    \dot\gamma^*_{ij} =\frac{\rho_\mathrm{a}\mf Q^2\dot\gamma_{ij}}{\eta }
    .
    \label{liquidnondim}
\end{equation}
Using \eqref{liquidnondim} and \eqref{model:air_nondim}, \eqref{noslipmassconsdim}-\eqref{noslipvertmomdim} give
\begin{equation}
    \frac{\partial {u}}{\partial{x}} + \frac{\partial{v}}{\partial{y}} = 0,\label{noslipmassconsnodim}
    \end{equation}
    \begin{equation}
    \mf^2\mR\left(\frac{\partial{u}}{\partial{t}}+{u}\frac{\partial{u}}{\partial{x}} + {v}\frac{\partial{u}}{\partial{y}}\right) = -\frac{\partial{p}}{\partial{x}} + \mf\frac{\partial{\tau}_{{x}{x}}}{\partial{x}} + \frac{\partial{\tau}_{{x}{y}}}{\partial{y}},\label{nosliphorizmomnodim}
    \end{equation}
    \begin{equation}
    \mf^4\mR\left(\frac{\partial{v}}{\partial{t}}+{u}\frac{\partial{v}}{\partial{x}} + {v}\frac{\partial{v}}{\partial{y}}\right) = -\frac{\partial{p}}{\partial{y}} + \mf^2\frac{\partial{\tau}_{xy}}{\partial{x}} + \mf\frac{\partial{\tau}_{yy}}{\partial{y}} - \mathcal{G},\label{noslipvertmomnodim}
\end{equation}
where 
\begin{equation}
    \mR = \frac{\rho_l\rho_\mathrm{a}Q^2}{\eta^2} \quad\mbox{and }\quad {\mG}=\frac{\rho_lgH^3}{\rho_\mathrm{a}Q^2}
\end{equation}
quantify the relative importance of liquid inertia to viscous stresses and of gravity to air inertia, respectively. We assume that terms of $O(\mf,\mf^2\mR)$ can be neglected,
but that the gravity parameter $\mathcal{G}=O(1)$. Hence, at leading order,
\begin{eqnarray}
    0 &=& -\frac{\partial{p}}{\partial{x}} + \frac{\partial{\tau}_{{x}{y}}}{\partial{y}},\label{nosliphorizmomLO}\\
    0 &=& -\frac{\partial{p}}{\partial{y}}  - {\mG},\label{noslipvertmomLO}
\end{eqnarray}
Similarly, inserting \eqref{liquidnondim} into the boundary conditions \eqref{noslipnormstressBCdim} and \eqref{noslipshearstressBCdim}, gives,
to leading order in $\mf$, 
\begin{equation}
    p =  P - \frac{1}{\mW}\frac{\partial^2h}{\partial x^2} ,\quad \tau_{xy} = {\tau}_\mathrm{w}\quad\mbox{at}\quad y = h,\label{noslipstressBCs}
\end{equation}
where the importance of air drag relative to surface tension is
gauged by the dimensionless group,
\begin{equation}
    \mathcal{W} = \frac{\rho_\mathrm{a}Q^2}{\mf^2\sigma H},
\end{equation}
which we assume is order one.
The constitutive law \eqref{stokesconstit} becomes
\begin{equation}
\begin{array}{ll}
   {\tau}_{ij} = 
       \left(1 + \frac{B}{{\dot\gamma}}\right){\dot\gamma}_{ij}  & \quad\mbox{if}\quad{\tau} \geq B \\
    {\dot{\gamma}}_{ij} = 0 & \quad \mbox{if}\quad{\tau} < B,
\end{array}
\label{nondimconstit}
\end{equation}
where
\begin{equation}
    B = \frac{\tau_\mathrm{y} H^2}{\rho_\mathrm{a}\mf Q^2}
\end{equation}
is a dimensionless parameter measuring the importance of the yield stress
in comparison to air drag; we (informally)
refer to $B$ as the Bingham number.

Combining \eqref{noslipvertmomLO} and \eqref{noslipstressBCs} gives
\begin{equation}
    p = \mathcal{G}(h-y) + P - \frac{1}{\mathcal{W}}\frac{\partial^2h}{\partial x^2},\label{noslipP}
\end{equation}
whilst \eqref{nosliphorizmomLO} and \eqref{noslipP} give
\begin{equation}
    \tau_{xy} = \frac{\partial p}{\partial x}(y-h) + \tau_\mathrm{w}.\label{nosliptauxy}
\end{equation}
Inserting the expressions \eqref{pxAIR} and \eqref{tauwAIR} into \eqref{noslipP} and \eqref{nosliptauxy} gives 
\begin{equation}
    \tau_{xy} = (y-h)\hat{G} + \frac{1}{(1-h)^2},\label{tauxyGS}
\end{equation}
where
\begin{equation}
    \hat{G}(x,t) = {\mathcal{G}}h_x-\frac{(h_x+2)}{(1-h)^3} - \frac{1}{\mathcal{W}}h_{xxx}.\label{GSdefn}
\end{equation}

At leading order, we expect the flow within the liquid layer to be composed of regions of shear-dominated flow, where the shear stress is asymptotically larger than the normal stresses and exceeds the yield stress, $|\tau_{xy}|>B$, and regions of plug-like flow where $|\tau_{xy}|\leq B$ and the normal stresses enter the leading-order balance. In regions of plug-like flow, the velocity profile is independent of $y$ at leading order, $u=u_{p}(x,t)$. Where $u_p$ varies with $x$, the fluid is a weakly yielded pseudo-plug in which the total stress exceeds the yield stress by an asymptotically small amount \citep{walton_axial_1991,BALMFORTH199965}. Since $\tau_{xy}$ is a linear function of $y$ \eqref{tauxyGS}, there are exactly two values of $y$ for which $|\tau_{xy}|=B$, which we call $y=\mathcal{Y}_\pm$, respectively. However, these locations may not lie within the fluid layer. To account for all the various configurations that are possible,
we define
\begin{equation}
  Y_\pm = \max[0,\min(h,\mathcal{Y}_\pm)],\label{Ypm}
\end{equation}
so that, within the layer, the fluid is fully yielded with shear-dominated flow for $0\leq y\leq Y_-$ and $Y_+\leq y \leq h$, and there is a region of plug-like flow in $Y_-<y<Y_+$. The switches in \eqref{Ypm} allow for the possibility
that one or more of these strata are not present in the flow.
A similar flow structure emerges in other thin viscoplastic film
flows \cite{hewitt_viscoplastic_2012,shemilt2023surfactant}. From \eqref{tauxyGS},
\begin{equation}
    \mathcal{Y}_{\pm} = h - \frac{1}{\hat{G}(1-h)^2} \pm \frac{B}{|\hat{G}|}.
    \label{Ypmdefn}
\end{equation}

The constitutive law \eqref{stokesconstit}, at leading order, implies that $\tau_{xy}=u_y + B\sgn(\tau_{xy})$ for $0\leq y \leq Y_-$ and $Y_+\leq y \leq h$, and $u_y=0$ for $Y_-<y<Y_+$. Combining this with \eqref{tauxyGS} and using the no-slip condition at $y=0$ then furnishes the vertical profile of the horizontal velocity. Once more, this depends on the detailed flow pattern and whether
or not one of the sheared layers or the pseudo-plug is present. However,
such details are taken care of in respecting the switches in \eqref{Ypm}, in which case the
flow profiles can all be summarized by the single formula,
\begin{equation}
    u = \left\{
    \begin{array}{cc}
    \vspace{8pt}
    \begin{array}{l}
      \half\hat{G}y(y-2h) \\
      \quad + y\left[(1-h)^{-2}+B\mathrm{sgn}(\hat{G})\right],
      \end{array}
           &  0<y\leq Y_-, \\
      \vspace{8pt}
      u_p ,
      &  Y_-<y\leq Y_+, \\
      \begin{array}{l}
         \half\hat{G}[y(y-2h) - Y_+(Y_+-2h)] + u_p \\
           \quad +  (y-Y_+)\left[(1-h)^{-2}-B\mathrm{sgn}(\hat{G})\right],
        \end{array}
      &  Y_+<y< h,
    \end{array}
    \right.\label{uBINGHAM}
\end{equation}
where 
\begin{equation}
u_p\equiv\half\hat{G}Y_-(Y_--2h) +  Y_-\left[(1-h)^{-2}+B\mathrm{sgn}(\hat{G})\right].
\end{equation}
Integrating \eqref{uBINGHAM} across the depth of the layer gives the horizontal flux,
\begin{multline}
    \hat{q} = -\frac{\hat{G}}{3}\left[h^3 + (h-Y_+)^3-(h-Y_-)^3\right] \\+ \frac{1}{2(1-h)^2}\left[h^2 + (h-Y_+)^2 - (h-Y_-)^2\right] \\
    + \frac{1}{2}B\sgn(\hat{G})[h^2 - (h-Y_+)^2 - (h-Y_-)^2].
    \label{qBINGHAM}
\end{multline}
From \eqref{noslipmassconsnodim} and the no-penetration condition at $y=0$,
we now arrive at the evolution equation,
\begin{equation}
    h_t + \hat{q}_x = 0 . \label{evoleqn}
\end{equation}

\subsection{Model equations}

We now introduce a minor rescaling to recast the equations of the reduced
model into a slightly more convinent form: we define
$\hat{x}=\mS x$ and $\hat{t}=\mS t$, where
\begin{equation}
    \mS = \mW^{1/3} = \left(\frac{\rho_\mathrm{a}Q^2}{\mf^2\sigma H}\right)^{1/3}.
\end{equation}
After substituting the rescaled variables, $\hat{x}$ and $\hat{t}$, into \eqref{GSdefn}-\eqref{evoleqn}, then dropping the hat decorations, we
then arrive at our final system of equations: the evolution equation
\begin{equation}
    h_t + q_x = 0,\label{evoleqn2}
\end{equation}
the horizontal flux,
\begin{multline}
    q = -\frac{{G}}{3}\left[h^3 + (h-Y_+)^3-(h-Y_-)^3\right] \\+ \frac{1}{2(1-h)^2}\left[h^2 + (h-Y_+)^2 - (h-Y_-)^2\right] \\
    + \frac{1}{2}B\sgn({G})[h^2 - (h-Y_+)^2 - (h-Y_-)^2],
    \label{qBINGHAM2}
\end{multline}
the rescaled pressure gradient,
\begin{equation}
  G = -\frac{(\mS h_x+2)}{(1-h)^3} - h_{xxx} + \mS{\mG} h_x,
  \label{eq:GG}
\end{equation}
and the yield-like surfaces,
\begin{equation}
    Y_\pm = \max\left[0,\min\left(h,\,h - \frac{1}{G(1-h)^2} \pm \frac{B}{|G|}\right)\right].\label{GSdefn2}
\end{equation}
We introduce another parameter, 
\begin{equation}
     \mathcal{J}\equiv B\mS^3 = \frac{\tau_\mathrm{y}H}{\sigma \mf^3}. 
\end{equation}
which is a plastocapillarity number, measuring yield stress relative to capillary
stresses \cite{jalaal_stoeber_balmforth_2021,van_der_kolk_tieman_jalaal_2023}. Unlike the Bingham
number, $B$, the plastocapillarity number,
$\mJ$, has no explicit dependence on the air speed. Therefore, where we look to examine the effect of varying the air speed while fixing the yield stress, relative to capillary stresses, we may fix the value of $\mJ$ while varying $\mS$. 

In the Newtonian limit, $B=\mJ=0$, the flux \eqref{qBINGHAM} becomes
\begin{equation}
    q = q_\mathrm{Newt} = -\frac{G}{3}h^3 + \frac{\mathcal{T}}{2}h^2,
    \label{qNewtonian}
\end{equation}
and the evolution equation becomes identical, when $\mathcal{G}=0$ and after some rescalings, to the one presented previously by Matar {et al. }\cite{matar2007interfacial}. Without the terms arising due to air-induced stresses in \eqref{qBINGHAM2}-\eqref{GSdefn2}, the viscoplastic evolution equation reduces to that appearing in other thin-film models of plastocapillarity \cite{jalaal_long_2016,jalaal_stoeber_balmforth_2021,van_der_kolk_tieman_jalaal_2023}. The first term on the right-hand side of \eqref{eq:GG} turns out to be critical for generating surface-wave instability in the current problem, and is qualitatively similar to the capillary pressure term that drives instability in viscoplastic linings of cylindrical tubes \cite{shemilt2022surface}.

\subsection{Boundary and initial conditions}

We compute solutions to \eqref{evoleqn}-\eqref{GSdefn2} in a domain,
$0\leq x\leq L$. We apply two different types of boundary and initial conditions. First, for the results presented in
\S\ref{sec:LSA}-\S\ref{sec:travel}, we assume that the domain is periodic,
and apply the initial condition,
\begin{equation}
    h(x,0) = \bar{h} + A\sin\left(\frac{2\pi x}{L}\right),
    \label{ICsin}
\end{equation}
where $A<\bar{h}<1$, and the perturbation amplitude $A$ is typically
taken to be $10^{-3}$. In this spatially periodic setting, we explore the
dynamics of linear instabilities, and 
take $L$ to be the wavelength of the most unstable mode.

Second, in \S\ref{sec:long}, we apply different boundary and initial conditions, focusing on an alternative spatial setting in which the domain is much longer. We do not consider a periodic domain, but fix the height and assume zero flux at the left boundary:
\begin{equation}
    h(0,t) = \bar{h},\quad q(0,t) = 0.\label{q0BC}
\end{equation}
At the right edge of the domain, we enforce
\begin{equation}
    h(L,t) = \bar{h},\quad h_x(L,t) = 0, \label{endBC}
\end{equation}
which allows for a non-zero flux out of the domain at $x=L$,
but that position is taken
sufficiently far downstream that, for practical purposes,
\eqref{endBC} has little impact on the solutions.
In this second spatial setting, we consider the evolution of finite-amplitude,
localised perturbations to the free-surface height, with
\begin{equation}
    h(x,0) = \left\{\begin{array}{cc}
        \bar{h} - F(x) &  \mbox{if} \quad x_0-1 < x < x_0,\\ 
        \bar{h} + F(x) &  \mbox{if} \quad x_0 < x < x_0+1,\\ 
        \bar{h} & \mbox{otherwise},
    \end{array}\right.
    \label{pertIC}
\end{equation}
where
\begin{equation}
     F(x) = A_b\left[1-\cos\left(2\pi(x-x_0)\right)\right]^2,
\end{equation}
and $A_b$ is a perturbation amplitude. We typically choose $x_0=3/2$; since we take $L$ to be large, this choice of $x_0$ localises the perturbation near the left edge of the domain. The form of the initial condition \eqref{pertIC} is chosen so that the initial mean height remains equal to $\bar{h}$. 


\subsection{Numerical methods}\label{numpty}

To ease computations, we regularise the Bingham constitutive law, replacing \eqref{nondimconstit} with
\begin{equation}
    \tau_{ij} = \left(1 + \frac{B}{|\dot{\gamma} + \delta|}\right)\dot{\gamma}_{ij},\label{constit_regGNF}
\end{equation}
where $\delta>0$ is a regularisation parameter. The modification
to the flux resulting from this regularisation is derived in \ref{app:reg}.
We generally take $\delta$ to be at most $10^{-4}$ in numerical simulations after checking that the exact value does not have a significant effect on the results.
To solve the regularised model as an initial value problem, we approximate the spatial derivatives using second-order centred finite differences on a grid of $N$ points, and solve the resulting system of equations using MATLAB's in-built
solver ode15s. We use at least $N=400$ grid points, and up to $N=10000$ when computing solutions in longer domains in \S\ref{sec:long}.

\section{Theoretical results}\label{sec:res_model}

\subsection{Instability of an almost flat layer}\label{sec:LSA}

The model equations admit a base state in which the fluid layer
has constant thickness $h=\bar{h}$. This state is given by
\begin{align}
  G &= -\frac{2}{(1-\bar{h})^3}, \\
  Y_- &= Y_0 \equiv \max\left\{0,\min\left[\bar{h},\,\half (1 + \bar{h})
    - \half B(1-\bar{h})^3\right]\right\},\\
    Y_+&=\bar{h}
\end{align}
together with a flux following from \eqref{qBINGHAM2}.
There are three distinct cases to consider based on the values of $\bar{h}$ and $B$: (i) if $B(1-\bar{h})^2<1$ then $Y_\pm=h$ in the base state and the whole layer is fully yielded; (ii) if $1+\bar{h}>B(1-\bar{h})^3>1-\bar{h}$ then $0<Y_-<h$, so there is a pseudo-plug adjacent to the free surface, but the fluid in $0<y<Y_-$ is fully yielded; (iii) if $B(1-\bar{h})^3>1+\bar{h}$ then $Y_-=0$, so the entire fluid layer is unyielded and motionless.

To determine the stability of such base states,
we set
\begin{equation}
    h = \bar{h} + Ae^{ikx+\lambda t}, \label{LSAIC}
\end{equation}
where $0<\bar{h}<1$ and linearize in the small amplitude $|A|\ll \bar{h}$.
Inserting \eqref{LSAIC} into \eqref{evoleqn2}-\eqref{GSdefn2}, and writing
$\lambda = \lambda_{r}+i\lambda_{i}$, gives a growth rate,
\begin{equation}
    \lambda_r = \frac{\bar{h}^3k^2V}{3}\left[\frac{\mS}{(1-\bar{h})^3} - \mS\mG - k^2\right],
    \label{LSA_omr}
\end{equation}
and phase speed,
\begin{equation}
   c=-\frac{\lambda_i}{k} = \left\{\begin{array}{lc}
   \frac{\bar{h}(1+\bar{h})}{(1-\bar{h})^4}-B, &
   B < \frac{1}{(1-\bar{h})^2}\\
   \frac{2\bar{h}^3V}{(1-\bar{h})^4} + \frac{Y_0(4\bar{h}-Y_0)}{(1-\bar{h})^3},
   & 
        \frac{1+\bar{h}}{(1-\bar{h})^3} > B \geq \frac{1}{(1-\bar{h})^2},
    \end{array}\right.\label{LSA_omi}
\end{equation}
where
\begin{equation}
    V = 1-\left(1-\frac{Y_0}{\bar{h}}\right)^3.
\end{equation}

\begin{figure}[t!]
    \centering
    \includegraphics[width=.9\linewidth]{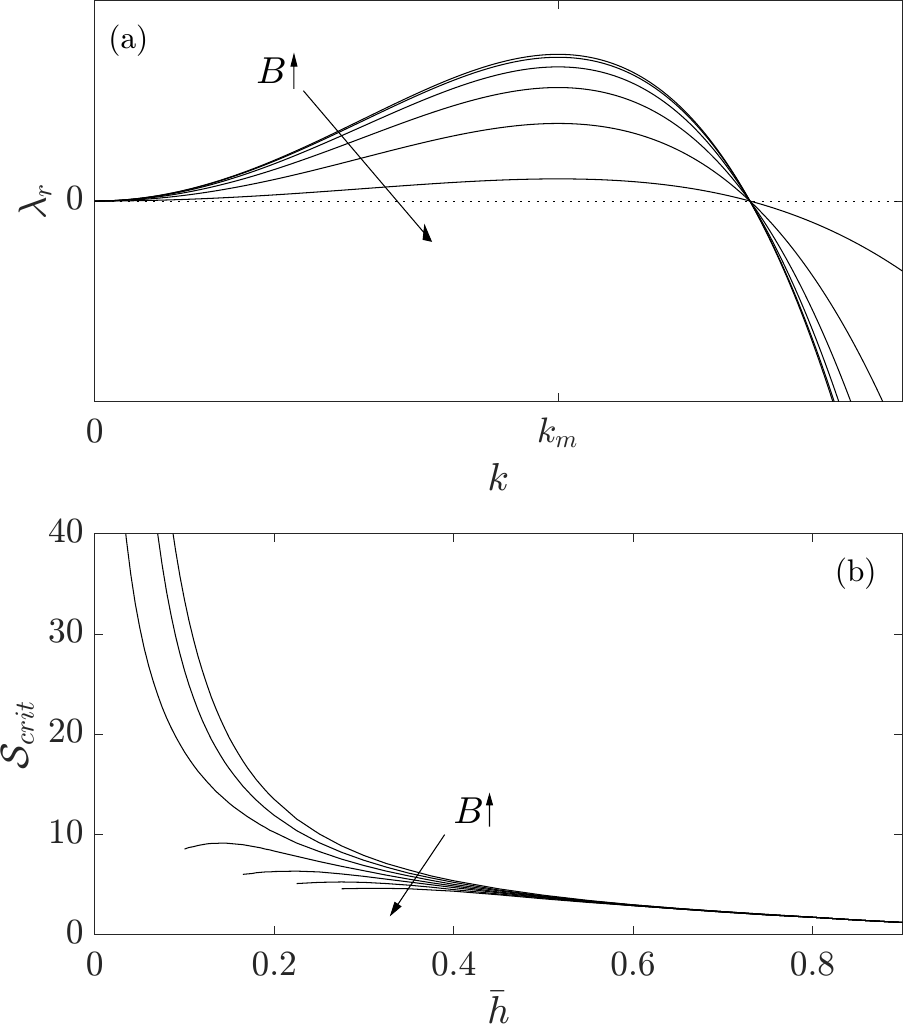}
    \caption{(a) Growth rate, $\lambda_r$, for $\mathcal{S}=10$, $\bar{h}=0.25$ and $B=\{0,2,2.2,2.4,2.6,2.8,3\}$. (b) Boundary between regions of convective and absolute instability for $B=\{0,0.5,1,1.5,2,2.5,3\}$. Instability is absolute if $\mS>\mS_\mathrm{crit}$ for given $B$ and $\bar{h}$. Where the lines terminate in (b) corresponds to the minimum $\bar{h}$ for which a uniform layer is in motion for the given $B$.}
    \label{fig:grBriggs}
\end{figure}

From \eqref{LSA_omr}, we see that there is instability if
$k^2<\mathcal{S}(1/(1-h_0)^3-\mathcal{G})$, and the most unstable
wavenumber is
\begin{equation}
  k_m = \sqrt{\frac{\mathcal{S}}{2}\left[\frac{1}{(1-\bar{h})^3}
      -\mathcal{G}\right]}.\label{LSAkm}
\end{equation}
In particular, without gravity, $\mathcal{G}=0$,
there is apparently no threshold in air speed (or, equivalently,
the air drag parameter $\mathcal{S}$) for long waves to be unstable.
However, \eqref{LSA_omr}-\eqref{LSAkm} do not account for the
yield criterion of the base state: if $B>(1+\bar{h})/(1-\bar{h})^3$
(in case (iii)),
that state is unyielded and no linear perturbation can break the fully
plugged layer to permit instability to develop. Moreover, $c=0$
for $B=(1+\bar{h})/(1-\bar{h})^3$.
In other words, the condition that the layer is linearly stable simply
corresponds to the yield threshold of the base state.
Also, in case (ii),
when $B>1/(1-\bar{h})^2$, increasing $B$ reduces the growth rate
but does not affect the most unstable wavelength (figure \ref{fig:grBriggs}a). 

From \eqref{LSA_omr} and \eqref{LSA_omi}, we can use the Briggs criterion \cite{briggs1964electron,huerre1990local} to predict whether the instability is absolute or convective for different values of $B$, $\bar{h}$ and $\mS$. 
Along the ray $x=\alpha t$, we can assume that there is some wavenumber, $k_\alpha$, such that 
\begin{equation}
    \frac{\partial\lambda}{\partial k}(k_\alpha) = i\alpha,\label{LSA:alpha}
\end{equation}
and the temporal growth rate along that ray is
\begin{equation}
    s_\alpha = \frak{Im}\left[i\lambda(k_\alpha) - \alpha k_\alpha\right]. \label{LSA:sk}
\end{equation}
Whether the system is absolutely unstable can be determined by the absolute growth rate, $s_0$, {\it{i.e.}} the temporal growth rate along the ray $x=0$ \cite{huerre1990local}. If $s_0>0$, then the growth rate at a fixed spatial position is positive, so the system is absolutely unstable. Figure \ref{fig:grBriggs}(b) shows the absolute/convective stability boundaries in $\bar{h}-\mS$ space for a range of values of $B$. As $B$ is increased, the critical value of $\mS$ above which there is absolute instability decreases, with this effect being more pronounced at lower $\bar{h}$. We can also identify rays other than $x=0$ along which the temporal growth rate takes certain values of interest. The rays along which the maximum growth rate is attained are $x-ct=\mathrm{const}$, where $c$ is given by \eqref{LSA_omi}. The ray, $x-\alpha_nt=\mathrm{const}$, along which there is marginal stability, {\it{i.e.}} the temporal growth is zero, is found by solving \eqref{LSA:alpha} and \eqref{LSA:sk} subject to $s_{\alpha_n}=0$. 


Note that 
there is a complication hidden in this long-wave formulation of the linear
instability problem: in case (ii), the model always assumes that the
pseudo-plug is yielded, even though the base state could be genuinely
unyielded there. The stability analysis therefore
implicitly assumes that the amplitude, $|A|$, is large enough that the perturbation breaks any such true plug,
allowing the free surface to deform and transforming that region
into the pseudo-plug \cite{balmforth2004roll}. 
We do not pursue this detail here,
assuming that the initial perturbation is sufficient to create
a pseudo-plug at the free surface so that the long-wave model
derived in \S\ref{sec:model_liquid} is valid.

\subsection{Nonlinear evolution in periodic domains}\label{sec:nonlinear_periodic}

\begin{figure}[t!]
    \centering
        \includegraphics[width = .49\textwidth]{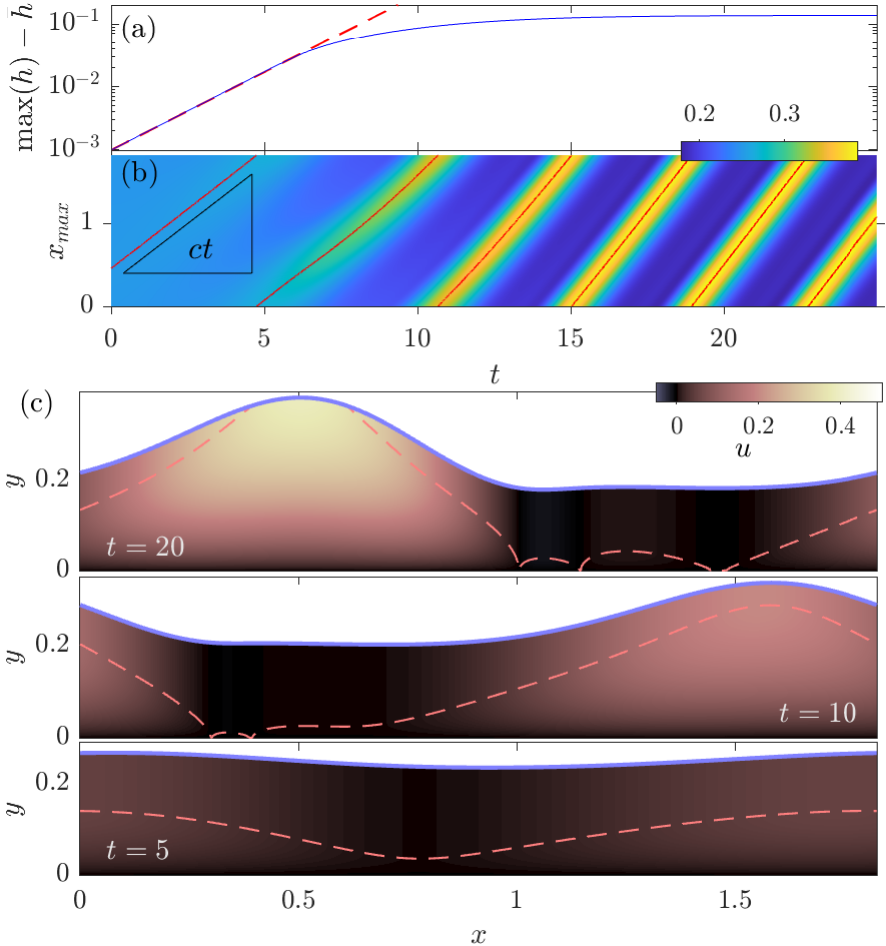}
    \caption{Numerical solution of the long-wave model for
      $\bar{h}=0.25$, $\mathcal{S}=10$, $\mJ=2500$, $L=2\pi/k_m$, $\mG=0$
      and initial conditions \eqref{ICsin}.
      (a) Time series of the deviation of the maximum layer height
      from the mean thickness.
      (b) Time series of $x_{\max}$, the location of the peak in $h$,
      superposed on a density plot of $h(x,t)$.
      The dashed red line and triangle
      indicate the linear growth rate \eqref{LSA_omr}
      and phase speed \eqref{LSA_omi}.
      (c) Snapshots,
      at the times indicated, of $h$ (blue), $Y_-$ (dashed red), and
      $u$ (as a density plot on the $(x,y)-$plane). In each snapshot,
      $Y_+=h$.
    }
    \label{fig:nonlinear_evolution1}
\end{figure}

\begin{figure}[t!]
    \centering
        \includegraphics[width = .49\textwidth]{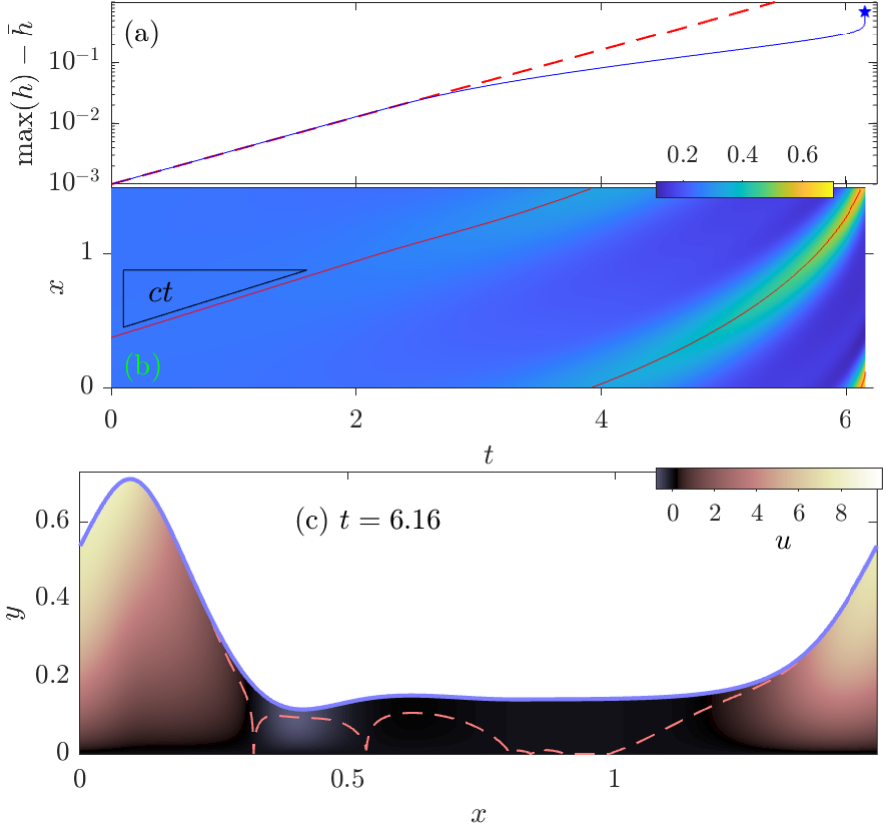}
    \caption{Numerical solution of the long-wave model for
      $\bar{h}=0.25$, $\mathcal{S}=15$, $\mJ=8438$, $L=2\pi/k_m$, $\mG=0$
      and initial conditions \eqref{ICsin}.
      (a) Time series of the deviation of the maximum layer height
      from the mean thickness.
      (b) Time series of $x_{\max}$, the location of the peak in $h$,
      superposed on a density plot of $h(x,t)$.
      The dashed red line and triangle
      indicate the linear growth rate \eqref{LSA_omr}
      and phase speed \eqref{LSA_omi}.
      The star in (a) indicates the time at which the computation
      ended due to the near blow-up of the solution.
      (c) Snapshot (at $t=6.16$)
      of $h$ (blue), $Y_-$ (dashed red), and
      $u$ (as a density plot on the $(x,y)-$plane);
      $Y_+=h$.
    }
    \label{fig:nonlinear_evolution2}
\end{figure}

To investigate the nonlinear evolution of liquid layers from an initially
near-flat configuration \eqref{ICsin} in a periodic domain, we turn to
numerical solutions of the long-wave model. For the most part and for brevity,
we neglect gravity ({\it i.e.} we set $\mG=0$). \ref{app:grav}
offers a deeper interrogation of gravitational effects. 
As stated in \S\ref{numpty}, we take
the domain length to be given by the most unstable wavelength
from the theory in \S\ref{sec:LSA}, $L=2\pi/k_m$, and trigger instability
using sinusoidal initial perturbations \eqref{ICsin} with amplitude $A=10^{-3}$.

Provided the initial layer is above the yield threshold, we observe two
regimes of behaviour. The first type of dynamics arises when
the air speed parameter $\mathcal{S}$ lies below a 
threshold that depends on the depth of the base state $\bar{h}$.
Figure \ref{fig:nonlinear_evolution1} shows a typical example
of the dynamics in this first regime:
the linear instability seeds the growth of a low-amplitude disturbance that follows the growth rate and wave speed predicted by 
\eqref{LSA_omr} and \eqref{LSA_omi}
(see figure \ref{fig:nonlinear_evolution1}a,b). At early times
(figure \ref{fig:nonlinear_evolution1}b, $t=5$), $0<Y_-<h$ everywhere, so
the base of the layer is fully yielded and a
pseudo-plug lies below the free surface.
Once the disturbance reaches
higher amplitude, however, nonlinearity arrests growth, leading to
the formation of a steadily propagating, nonlinear wave
(figure \ref{fig:nonlinear_evolution1}c, $t=20$).
Over the main body of the wave,
the fluid is strongly yielded, with
$Y_-=h$ at the wave's peak, implying that the entire
layer is fully yielded there. Over a shallower region between
the wave crests, fluid remains more weakly yielded, with $Y_-$ close to zero. In this region,
the free surface undulates and $Y_-$ forms bumps, with the direction of
flow changing between each of these bumps. Similar capillary undulations
are observed in various other viscoplastic thin-film flows with surface
tension \citep{jalaal_long_2016,jalaal_stoeber_balmforth_2021,shemilt2025viscoplasticity}.

\begin{figure}[t!]
    \centering
    \includegraphics[width=\linewidth]{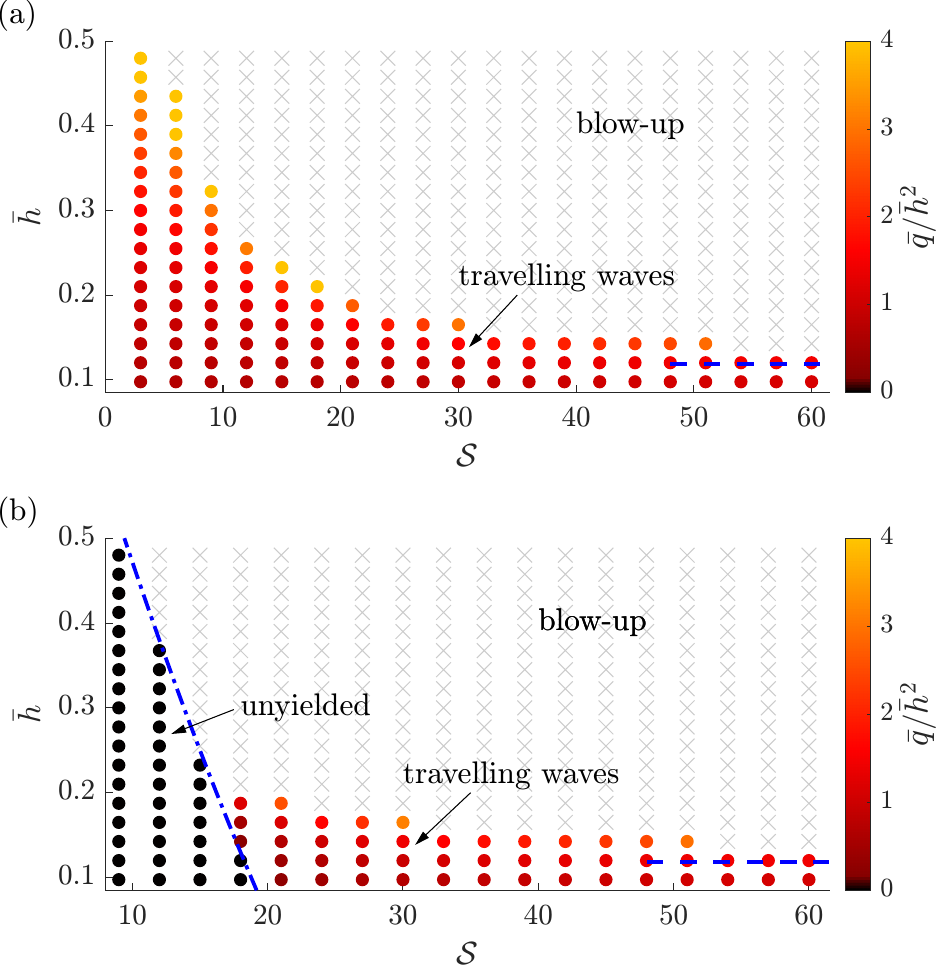}
    \caption{Data from numerical computations with (a) $\mathcal{J}=0$ and (b) $\mathcal{J}=10^4$, all with $\mG=0$ and $L=2\pi/k_m$. Each point represents one computation. Colour indicates the mean horizontal liquid flux, $\bar{q}$, scaled by $\bar{h}^2$, at the end time of the simulation. Grey crosses indicate solutions that were stopped prior to finite-time blow-up. Otherwise, computations were run to a maximum time of $150/\lambda_r$, where $\lambda_r$ is the Newtonian growth rate \eqref{LSA_omr}. The dot-dashed line in (b) shows the yielding
      threshold for the base state, $\mS^3(1+\bar{h})=\mJ(1-\bar{h})^3$. The dashed line shows the
      large-$\mathcal{S}$ prediction for the critical $\bar{h}$ for blow-up
      (see \ref{app:largeS}). }
    \label{fig:carp}
\end{figure} 

In the second regime of behaviour, for higher $\mathcal{S}$,
the initial phase of
evolution remains similar, 
as illustrated by a second solution shown in figure
\ref{fig:nonlinear_evolution2}.
However, instead of nonlinearity arresting the
instability at later times, growth suddenly accelerates
with the layer height abruptly increasing towards $h=1$
(the roof of the channel). At this stage, computations grind to a halt
with time steps becoming excessively short. We interpret this
halt to signify a blow-up of the solution in finite time,
although the computations cannot truly confirm this singular
behaviour. The snapshot of the solution shown in figure
\ref{fig:nonlinear_evolution2}(c) highlights how a
large fraction of the fluid layer
has yielded en route to blow-up.

The division of the dynamical behaviour into two regimes
is illustrated further in 
figure \ref{fig:carp}, which shows results from suites of computations
with varying $\bar{h}$ and $\mathcal{S}$.
The initial-value problems
shown in this figure are marked with a filled circle when instability
saturates into a nonlinear wave; the colour indicates the
corresponding mean fluid flux $\bar{q}$. On the other hand,
when the solution appears to blow up in finite
time, the computation is marked by a grey cross.

Figure \ref{fig:carp}(a) 
corresponds to the Newtonian case, $\mathcal{J}=0$, and highlights how
the split into the two regimes arises
whether or not the fluid has a yield stress.
For a Newtonian liquid with $\mS\gg1$, we can make an asymptotic
prediction for the critical depth $\bar{h}$ above which blow-up occurs;
see \ref{app:largeS}. In this limit,
we find that the shape of a travelling wave is determined by a balance
between inertial effects and surface tension, and that solutions
exist only for $\bar{h}<\bar{h}_\mathrm{c}\approx0.119$. The
division between the two regimes for $\mS\gg1$ in figure \ref{fig:carp}a
matches well with $\bar{h}_\mathrm{c}$.

Figure \ref{fig:carp}(b) shows results
for viscoplastic fluid with $\mathcal{J}=10^4$. In our model,
fixing $\mJ$ while varying $\mS$
corresponds to prescribing the yield stress and surface tension of the
liquid then varying the air speed.
Notably, in figure \ref{fig:carp}(b), the two dynamical
regimes are interrupted by the yielding threshold at the lowest
values of $\mathcal{S}$. 
As discussed in \S\ref{sec:LSA}, the threshold for a uniform layer to be unyielded is $(1+\bar{h})<B(1-\bar{h})^3$ or, in terms of $\mJ$, $\mS^3(1+\bar{h})<\mJ(1-\bar{h})^3$.
As seen in figure \ref{fig:carp}(b), the yielding threshold
actually eliminates
the nonlinear wave regime entirely for larger layer depths.
Thus, when $\bar{h}$ is relatively large, if the air speed (or $\mS$)  is
increased gradually from a low value, there can be a sudden transition
from no motion to significant motion and blow-up. By contrast, for a
Newtonian liquid, nonlinear waves always precede blow-up and
there is a gradual increase in mean flux $\bar{q}$ as $\mS$ is raised
(figure \ref{fig:carp}a).

For larger $\mS$, the critical $\bar{h}$ below which 
travelling waves form
for a viscoplastic layer is similar to that for a Newtonian
layer, approaching the
same limiting value for $\mS\to\infty$.
This feature arises naturally because,
when $\mS$ is increased with fixed $\mJ$, the original yield stress
parameter $B$ decreases rapidly, and so the layer is expected to become
almost fully yielded and behave like a Newtonian layer.
Closer to the yielding threshold, the yield stress has a greater impact
on the wave dynamics, as we explore
in more detail below.

\subsection{Steady periodic waves}\label{sec:travel}

\begin{figure*}[t!]
    \centering
    \includegraphics[width=.8\linewidth]{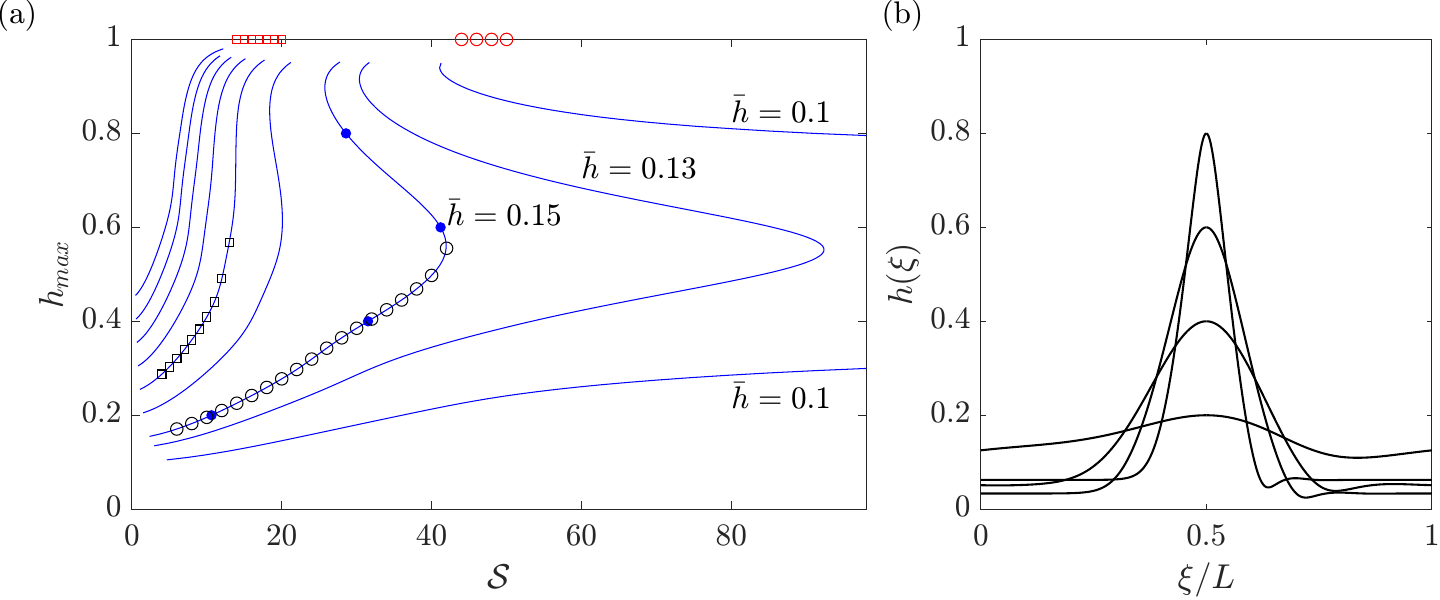}
    \caption{(a) Maximum height, $h_{max}$, of travelling-wave solutions. Each line corresponds to a different $\bar{h}\in\{0.1,0.13,0.15,0.2,0.25,0.3,0.35,0.4,0.45\}$. All solutions have $\mG=0$ and $L=2\pi/k_m$. Final maximum height also shown from solutions to the initial value problem (\ref{evoleqn}, \ref{qGNF}) with \eqref{ICsin} for $\bar{h}=0.15$ (circles) and $\bar{h}=0.25$ (squares). Red circles/squares indicate values of $\mathcal{S}$ for which finite-time blow-up occurred. (b) Example travelling-wave solutions with $\bar{h}=0.15$, corresponding to the solid blue dots in (a). }
    \label{fig:Ntravel}
\end{figure*}

The model equations admit steady nonlinear wave solutions for
which $h=h(\xi)$ and $q=q(\xi)$, where $\xi=x-Ut$ is a 
travelling-wave coordinate and $U$ is the nonlinear wavespeed.
For these solutions, an integral of \eqref{evoleqn2} furnishes
\begin{equation}
    q = C+Uh,\label{traveleqn}
\end{equation}
for some integration constant $C$. In combination
with the flux law, which relates $q$ to $h$ and its derivatives
{\it via} $G$, we then arrive at
a third-order ordinary differential equation (ODE) for 
the wave profile $h(\xi)$. In practice, we use
the regularised version of the flux relation in \eqref{qGNF}.
After imposing periodic boundary conditions, 
the mass conservation constraint,
\begin{equation}
\frac{1}{L} \int_0^L h(\xi) \; {\rm d}\xi = \bar{h},
\end{equation}
and a condition that eliminates translational invariance
(such as $h_\xi(0)=0$), the task is then to solve the ODE for $h(\xi)$
and determine the two parameters $U$ and $C$ as part of the solution.
The in-built function bvp4c in Matlab suffices for this purpose,
with initial guesses provided either from numerical
solutions of the initial-value problem, or from continuation
from other parameter settings.
Again we take the domain length to be $L=2\pi/k_m$.


\subsubsection{Newtonian waves}

First, we interrogate the structure of Newtonian travelling waves.
Sample solutions are displayed in figure \ref{fig:Ntravel}.
Here, we fix $\bar{h}$ and track the travelling-wave solutions
with varying $\mS$; solution branches with several mean depths are shown.
When $\bar{h}\gtrsim0.25$, $\max(h)$ increases monotonically with $\mS$
(figure \ref{fig:Ntravel}a). For lower mean depths
($\bar{h}=0.13,0.15,0.2$ in figure \ref{fig:Ntravel}a), however,
the solution branches ascend non-monotonically,
turning back to smaller $\mS$ when wave heights reach values around a half,
before returning to higher $\mS$ once the peak
nears the channel roof. Somewhere between $\bar{h}=0.13$ and $\bar{h}=0.1$,
the first turn-back apparently diverges to the limit $\mS\to\infty$,
breaking the solution branch
with the smallest mean depth of $\bar{h}=0.1$ into two disconnected pieces.

The structure of the steady travelling-wave branches
evident in figure \ref{fig:Ntravel}(a) connects with the outcome
of Newtonian initial-value computations:
for the thinner layers with $\bar{h}=0.13,0.15,0.2$,
blow-up occurs when $\mS$ exceeds the first turn back
(as seen in figure \ref{fig:Ntravel}a,
which also includes data from a suite of initial-value computations
with $\bar{h}=0.15$).
For a thicker layer with $\bar{h}=0.25$, there is no saddle node
at which the solution branch turns back. Nevertheless,
the initial-value problem 
blows up in finite time beyond a value of $\mS$ that
coincides with where the corresponding travelling-wave
solution branch begins to rise steeply
({\it cf.} figure \ref{fig:Ntravel}a).
Evidently, the sudden rise in the branch places the nonlinear
wave in an inaccessible part of phase space; instead, the
initial-value problem becomes launched towards blow-up.
For the lowest mean depth with $\bar{h}=0.1$, the
steady wave branch continues with a relatively shallow gradient
and no interruption all the way
to the limit $\mS\to\infty$. Consequently, in the initial-value
problem linear instabilities always saturate into nonlinear
waves and no blow-up occurs.

\subsubsection{Viscoplastic waves}

\begin{figure*}[t!]
    \centering
    \includegraphics[width=.9\linewidth]{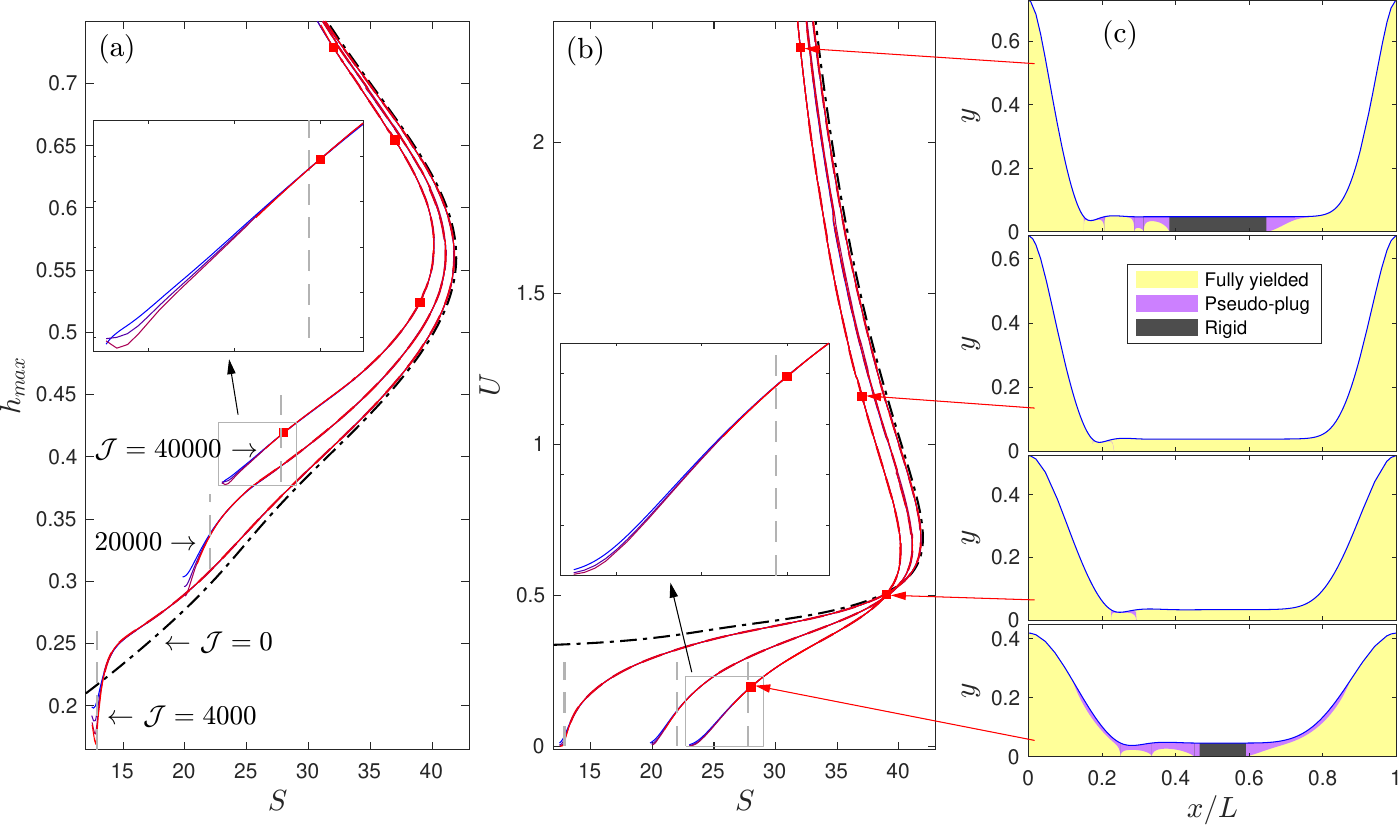}
    \caption{(a) Maximum height $h_{max}$ and (b) wave speed $U$
      of travelling-wave solutions with
      $\mJ=\{4000,20000,40000\}$ (red/blue lines), and $\mJ=0$
      (dot-dashed black), all with  $\bar{h}=0.15$. 
      All solutions have $\mG=0$ and $L=2\pi/k_m$.
      For each $\mJ$, the
      different coloured lines correspond to solutions with regularisation
      parameters of $\delta=10^{-j/2}$ for $j=5,6,7,8$ (color-coded from red
      to blue); also shown are solutions with $\delta=10^{-8}$ (dashed red)
      which can only be computed when no true plugs spans the layer.
      The difference between the solutions with different values
      of $\delta$ is only visible near the termination of the solution branches
      for small $h_{max}$, as illustrated further by the magnifications
      for the case with $\mJ=40000$ shown in the insets.
      The vertical dashed lines cutting through the three sets
      of solution branches indicate the yielding threshold of the
      corresponding base state.
      (c) Example travelling-wave solutions with $\mJ=40000$, corresponding
      to the red squares in (a) and (b). The colours
      indicates where the fluid is fully-yielded (yellow), with $y<Y_-$
      and $y>Y_+$, where there are pseudo-plug (purple),
      with $Y_-<y<Y_+$, or where the layer is rigid (black),
      with $Y_-=0$ and $Y_+=h$. }
    \label{fig:VPtravelnj}
\end{figure*}

Figure \ref{fig:VPtravelnj} illustrates the impact of the yield stress
on travelling-wave structure, for the solution branch with $\bar{h}=0.15$.
Cases with three values of $\mJ$ are presented. For each case,
the branch still turns back for peak amplitudes of around a half when
$\mS$ reaches values near 40. Indeed, beyond that saddle node the branches
all lie close to their Newtonian relative, indicating that the yield
stress does not play a strong role. 
This feature is reinforced
from a closer inspection of the wave structure
in the vicinity of the saddle node, which shows that the layer is almost
fully yielded here (see the sample solutions shown in figure
\ref{fig:VPtravelnj}c). As one ascends further up the branch, however,
the situation becomes less clear: once fluid becomes collected into the
prominent peaks of the wave, the layer becomes relatively thin
in between. This thinning permits the yield stress to take effect
again, plugging up that region.

More significant impacts of the yield stress are visible
below the saddle node, where the solution branches approach the yielding
threshold for the original base state (indicated by the
vertical dashed lines in figure \ref{fig:VPtravelnj}a,b).
Here, the nonlinear wavespeed decreases to zero at some critical value
of $\mS$. Nevertheless, the ultimate fate
of the solution branches is less clear because the regularisation 
of the constitutive law becomes significant
at this stage (branches with
several values of $\delta$ are shown in figure \ref{fig:VPtravelnj}a,b;
these are indistinguishable in the figure except at the lowest
wavespeeds); we terminate the branches before
the regularisation becomes excessive (beyond the termination points,
the nonlinear waves creep forward at speeds controlled by $\delta$).
Despite this, it is clear that
the solution branches all descend past the yielding threshold
of the base state, and the peaks of the waves remain higher
than $\bar{h}$. In other words, the liquid layer rigidifies with a
residual stationary wave of finite amplitude.

\begin{figure*}[t!]
        \centering
        \includegraphics[width = .85\textwidth]{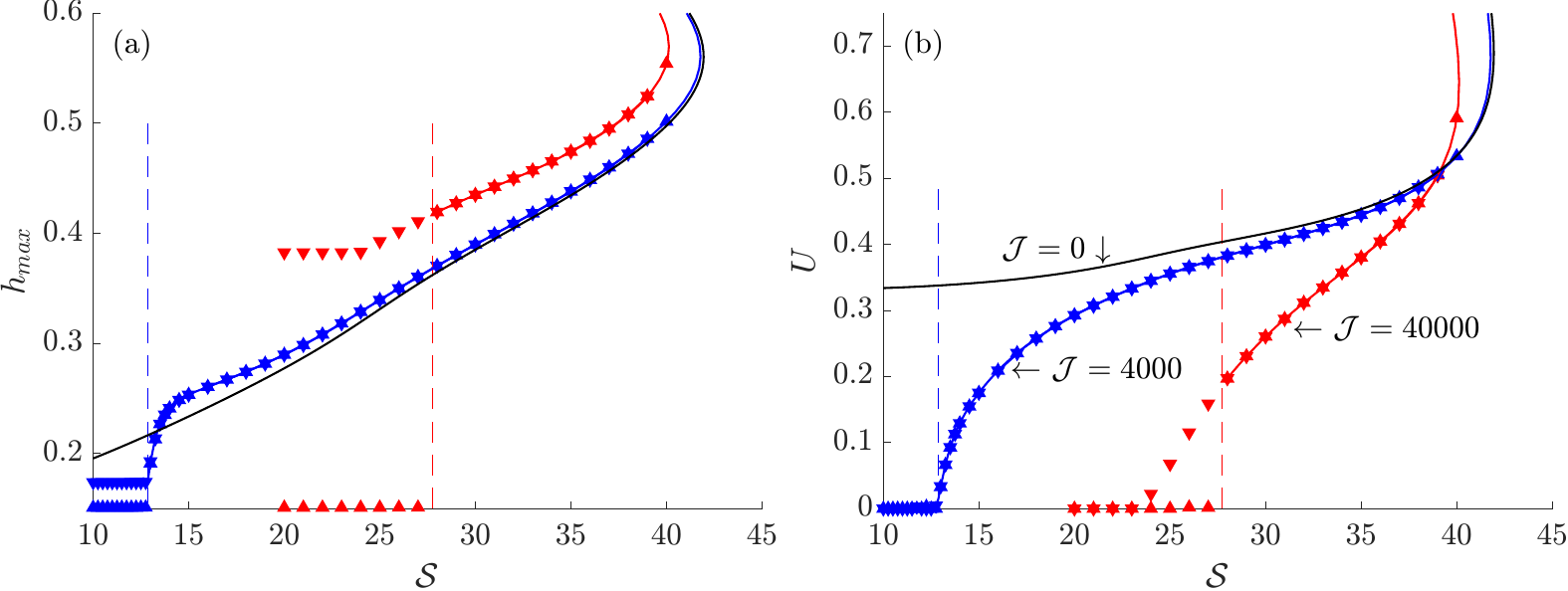}
        \caption{(a) Maximum height $h_{max}$ and (b) wave speed $U$ of
          travelling-wave solutions with $\mJ=40000$ (red), $\mJ=4000$ (blue)
          and $\mJ=0$ (black), all with $\bar{h}=0.15$, $\mG=0$ and
          $L=2\pi/k_m$. The solid lines denote solutions to \eqref{traveleqn}
          with $\delta=10^{-4}$ ({\it cf.} figure \ref{fig:VPtravelnj}).
          Triangles are late-time values from suites of initial-value computations, where the value of $\mS$ was incrementally increased then decreased. For each computation, the final solution for $h$ was used as the initial condition for the next value of $\mS$. Upwards pointing triangles denote the upwards sweep in $\mS$, and downwards pointing triangles denote the downwards sweep. Simulations were run until $h_{max}$ converged to its final value. 
    }
    \label{fig:VPtraveljs}
\end{figure*}

The structure of the viscoplastic wave branches
implies that the transition to instability must become discontinuous and 
hysteretic with a yield stress: if the air speed (or $\mS$)
above a flat layer is gradually raised above the yielding threshold, then
the onset of motion immediately leads a nonlinear wave with a finite
amplitude. That amplitude is dictated by where the vertical dashed line
in figure \ref{fig:VPtravelnj} (denoting the yielding threshold)
cuts through the relevant solution branch. The wave amplitude strengthens
as the air speed is increased further. However, if the air speed is then
reduced, the wave adjusts to follow the solution branch to lower $\mS$,
eventually reaching the termination point below the yielding threshold,
where the layer becomes unyielded but not flat.
The discontinuous, hysteretic transition is illustrated
in figure \ref{fig:VPtraveljs}
for the two cases, $\mJ=40000$ and $4000$. Here, the results of suites
of initial-value problems are presented, first increasing the air speed
up through the yielding threshold to near the saddle node
(upward directed triangles), then
decreasing it back to the residual rigidified wave (downward directed
triangles).

\subsection{Transient dynamics in longer domains}\label{sec:long}

Computations in spatially periodic domains with length $2\pi/k_m$
permit one to explore the nonlinear dynamics of the most unstable waves in 
a relatively simple setting. However, in longer spatial domains with 
finite ends, the dynamics may be different
as a result of the sweeping action of the overlying air flow
and interactions between multiple waves.
To explore this alternative scenario, we consider the second set of
boundary and initial conditions in \eqref{q0BC}-\eqref{pertIC}.
In addition, motivated by the experiments of \S\ref{sec:expts}, we focus
primarily on relatively large values of $\mS$
(which characterize those experiments).

\begin{figure*}[t!]
    \centering
    \includegraphics[width=\linewidth]{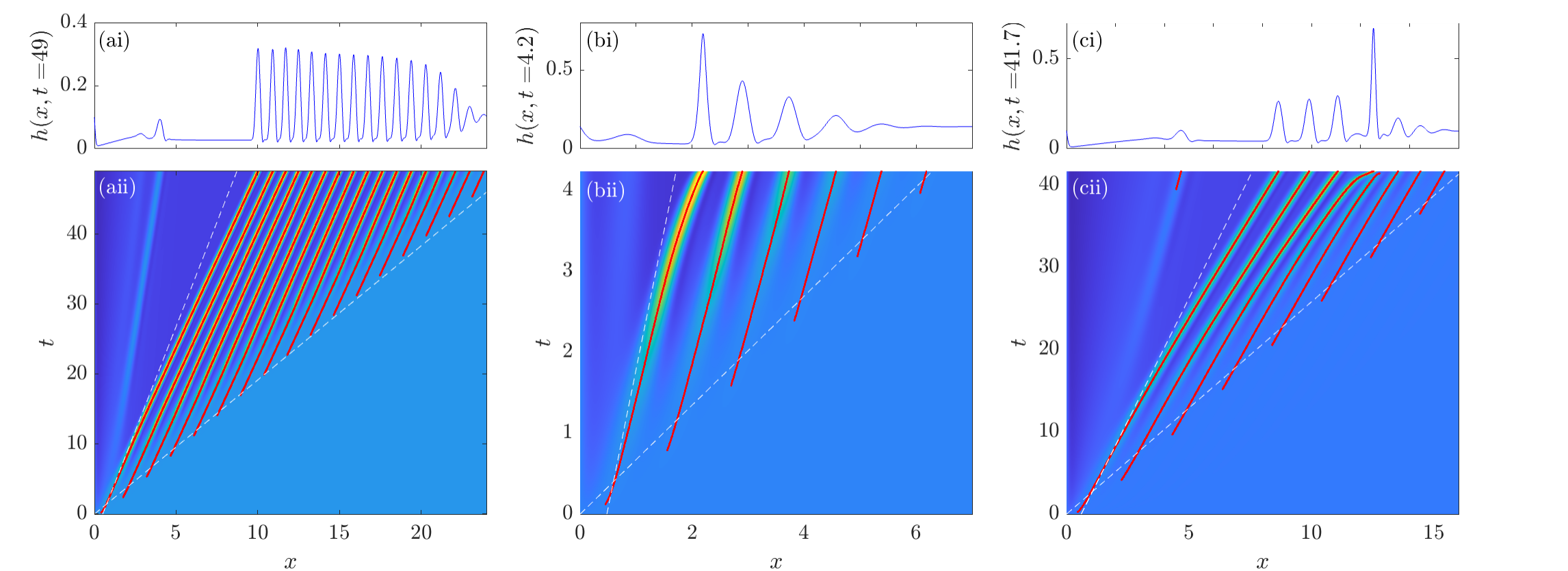}
    \caption{ Newtonian solutions ($\mJ=0$) in long domains with
      boundary conditions \eqref{q0BC}-\eqref{endBC} and initial conditions
      $h(x,0) = \bar{h}$, where
      (a) $\bar{h}=0.1$, $\mathcal{S}=55$,
      (b) $\bar{h}=0.14$, $\mathcal{S}=55$, and
      (c) $\bar{h}=0.1$, $\mathcal{S}=40$. Top panels (i) show
      snapshots shortly before the end of each computation
      (corresponding to the wave packet reaching the right end of the domain
      in (a), and blow-up for (b) and (c));
      lower panels (ii) display density plots of $h(x,t)$. In (ii), the red lines are the locations of wave peaks, determined as local maxima where $h(x,t)\geq1.01\bar{h}$. There are two dashed white lines in (ii): $x-ct=\pi/k_m$, a ray along which the temporal growth rate is maximal, where $c$ is given by \eqref{LSA_omi}, and $x-\alpha_nt=\pi/k_m$, along which the temporal growth rate is zero, {\it{i.e.}} there is a solution to \eqref{LSA:alpha} and \eqref{LSA:sk} with $\alpha = \alpha_n$ and $s_{\alpha_n}=0$. 
    }
    \label{fig:Nlong}
\end{figure*}

\subsubsection{Newtonian dynamics}

Again, we first consider the Newtonian problem. When the layer is initially
flat ($A_b=0$), waves do not immediately form, but the air flow sweeps fluid away
from the left end of the domain, leaving a trough near $x=0$ that is not
replenished because of the zero-flux boundary condition; see
figure \ref{fig:Nlong}, which shows three sample initial-value computations.
The trough constitutes a natural perturbation to the free surface,
which then triggers the formation of a first wave near $x=0$.
As this wave then propagates to the right,
more waves become triggered in front, creating a moving wave packet.
In the example shown in figure \ref{fig:Nlong}(a), for $\bar{h}=0.1$,
the wave packet widens into an approximately periodic wave train, with new waves forming at the front edge of the wave packet as it propagates. 
In such a case, we can expect the dynamics of these waves to resemble,
at least qualitatively, the dynamics of the periodic waves discussed
in the previous sections. 

The location of the front edge of the wave packet is well-approximated by, $x=\alpha_nt$, the ray along which the temporal linear growth rate is zero (see the discussion in \S\ref{sec:LSA}). The back edge of the wave packet is bounded by $x-ct=\pi/k_m$, a ray along which the linear temporal growth is maximised. When waves are forming and have low amplitude, their wave speed is close to the linear wave speed, $c$, but the waves quickly grow and accelerate to attain a nonlinear wave speed that is faster than $c$ (figure \ref{fig:Nlong}). 
In its wake, the wave packet leaves an almost uniform, thinned layer.
However, the trough near $x=0$ continues to deepen, shedding
a few smaller waves at later times. The values of $\mS$ and $\bar{h}$ in figure \ref{fig:Nlong}(a) are such that the Briggs criterion would predict absolute instability (figure \ref{fig:grBriggs}b), but this criterion has limited relevance here since the continual thinning of the layer near $x=0$ means that waves do not necessarily continue to form at any fixed spatial location. 

For a deeper initial layer, the growing disturbances within
the widening packet are unable to saturate into nearly steady nonlinear
waves. Instead, blow-up occurs, first for the wave at the rear of the
packet; see the example in figure \ref{fig:Nlong}(b) with $\bar{h}=0.14$.
For $\mS\gg1$, blow-up is expected to occur for periodic waves when
$\bar{h}\gtrsim0.119$ (\ref{app:largeS}), which is indeed the case
for the initial depth of the solution in figure \ref{fig:Nlong}(b).

The third example in figure \ref{fig:Nlong} shows a computation in which
the initial depth again lies below the critical value for blow-up
in a periodic domain. In this example, the wave packet and its
nearly regular train develops initially as in the case
shown in figure \ref{fig:Nlong}(a) (with a larger value of $\mS$).
This time, however, the wave train does not maintain its even spacing,
with two of the component waves interacting more strongly. The interaction
leads to a coalescence of the two waves, which sharply increases the amplitude
of the combined wave and then triggers blow-up.
Complex wave interactions have been noted previously in related models of
Newtonian two-layer flow \citep[e.g., ][]{matar2007interfacial}.

\subsubsection{Effects of a yield stress}\label{sec:longY}

\begin{figure}[t!]
    \centering
    \includegraphics[width=.9\linewidth]{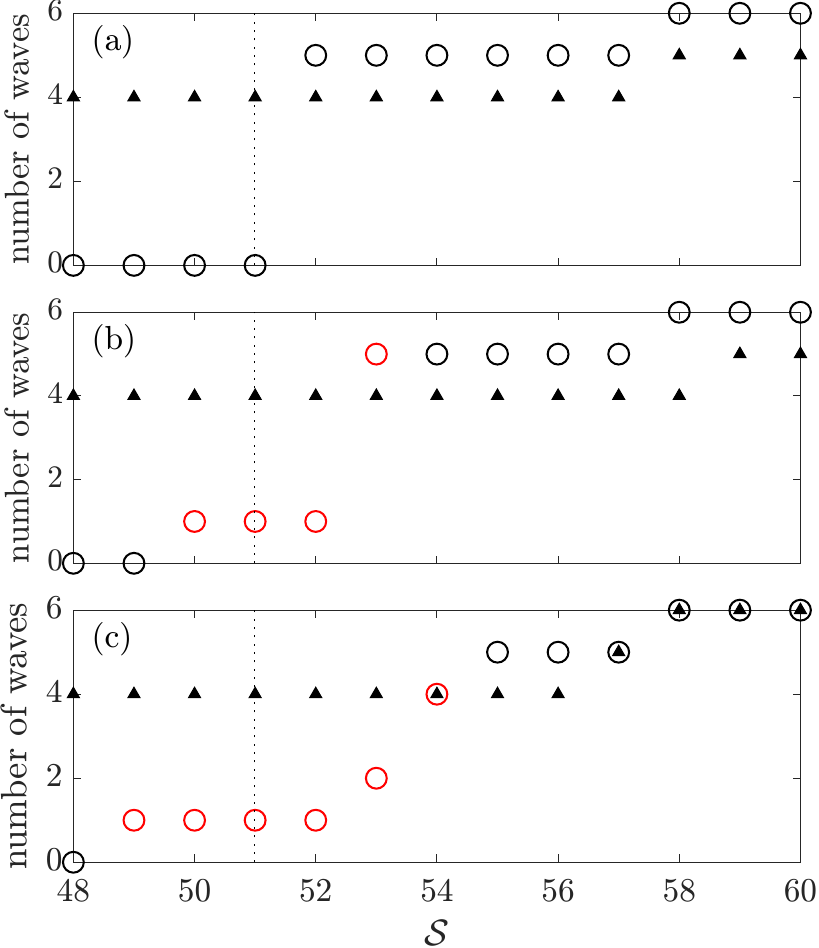}
    \caption{Maximum number of waves with peak height at least $0.14$
      observed in computations with varying $\mS$, initial mean layer depth $\bar{h}=0.1$, domain length $L=10$
      and different perturbation amplitudes: (a) $A_b=0$, (b) $A_b=0.01$,
      (c) $A_b=0.02$. The circles denote viscoplastic layers with
      $\mJ=2\times10^5$; the triangles are Newtonian. Computations are
      stopped when $h(L-2,t)$ deviates from $\bar{h}$ by more than $5\%$
      (black symbols), or when finite-time blow-up occurs (red symbols).
      The number of waves at any given time is calculated as the number of
      local maxima with $h>0.14$, and we plot the maximum number observed at
      any time during the computation. 
    }
    \label{fig:number_waves}
\end{figure}

For a viscoplastic layer, a first difference with the Newtonian dynamics
is the yield threshold: if the layer is initially flat, no motion
arises below the corresponding critical air speed. The trough
which triggers waves of instability in figure \ref{fig:Nlong}
cannot then form, and the layer remains static. However,
above the yielding threshold, a trough is able to form.
The model predicts dynamics that are similar to those seen in
the Newtonian case in figure \ref{fig:Nlong},
although, as discussed in \S\ref{sec:travel}, wave speeds and
amplitudes become modified by viscoplasticity. This is
illustrated in figure \ref{fig:number_waves}(a),
which compares the outcome of suite of computations for $\bar{h}=0.1$
and varying air speed ($\mS$) for both Newtonian
($\mJ=0$) and viscoplastic ($\mJ=2\times10^5$) layers.
For these computations, blow-up is not observed, and a quasi-steady
wave packet develops that eventually reaches the right end of the
domain. Plotted is the maximum number of waves contained in the packet.

\begin{figure*}[t!]
    \centering
    \includegraphics[width=0.8\linewidth]{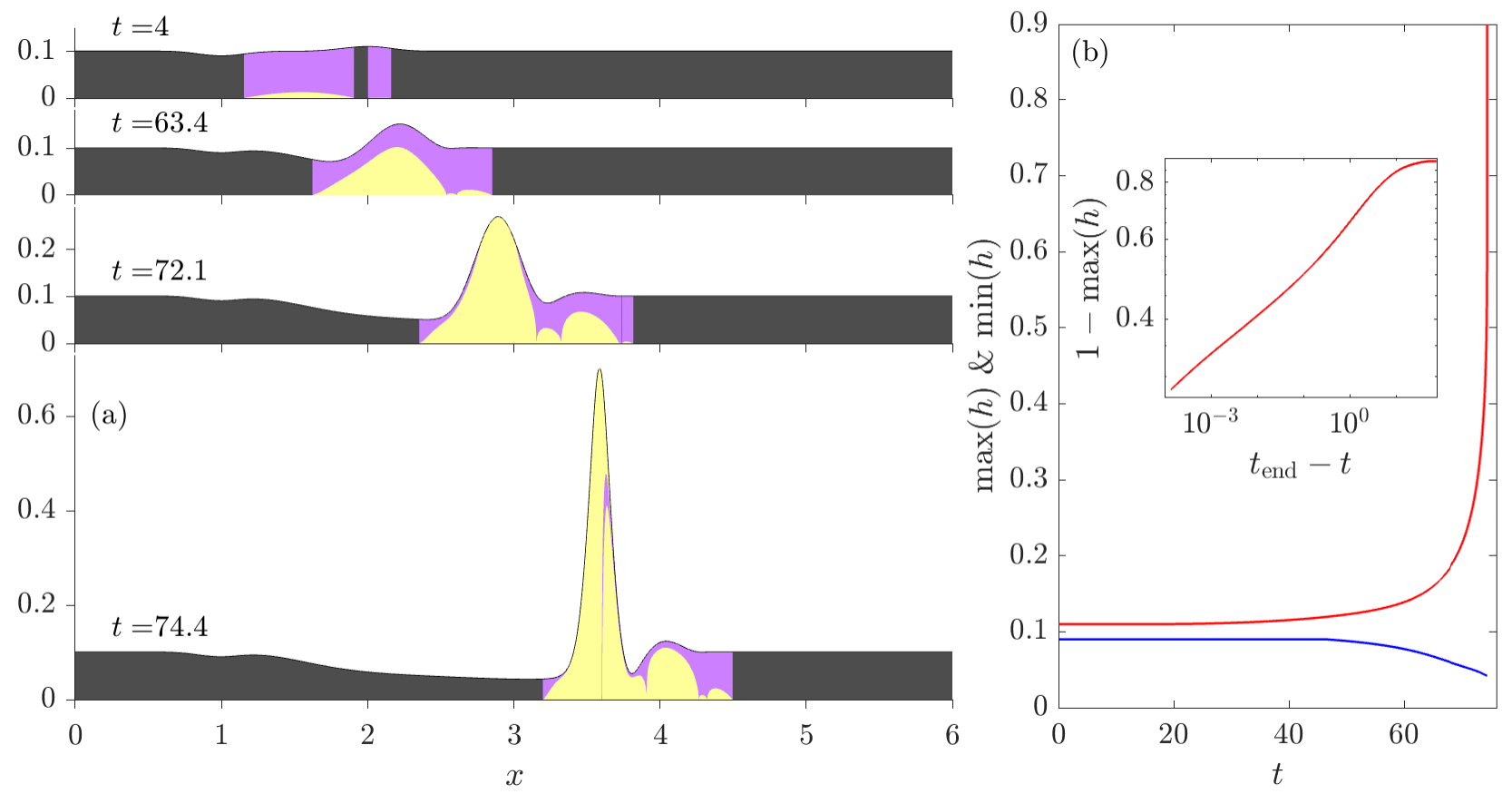}
    \caption{Solution to the evolution equation \eqref{evoleqn2} with \eqref{qGNF}, with initial conditions \eqref{pertIC}, $h_c=0.1$, $\mathcal{J}=2\times10^5$, $\mathcal{S}=50$, $A_b=0.01$. 
    (a) Snapshots showing the layer shape, with fully-yielded fluid (yellow), pseudo-plug (purple) and rigid regions (black) indicated, as determined by the values of $Y_\pm$. (b) Time evolution of the maximum (red) and minimum (blue) heights. Inset shows the difference between the maximum height and $1$, where $t_\mathrm{end}$ is the time at which the simulation was stopped due to finite-time blow-up. }
    \label{fig:longVPeg}
\end{figure*}

More interesting is that a 
qualitatively different behaviour emerges for viscoplastic layers
that are mostly below the yield threshold, but activated by
a localised, finite-amplitude perturbation to the free surface.
That is, for the initial condition \eqref{pertIC} with $A_b>0$.
In such cases, there is a range of values of $\mS$, for a given $\mJ$,
such that the fluid will yield locally around the site of the perturbation,
but remain unyielded elsewhere. One such example is shown in 
figure \ref{fig:longVPeg}. The local yielding is evident in
the early time snapshot at $t=4$ in figure \ref{fig:longVPeg}(a),
and triggers the formation of a wave from the site of the perturbation
(see the snapshot at $t=63.4$). The wave then
grows and accelerates as it travels through the domain,
depositing in its wake a much shallower unyielded layer.
The height of the wave increases rapidly (figure \ref{fig:longVPeg}a,b),
which eventually leads to finite-time blow-up.
Because the fluid ahead of the oncoming wave remains unyielded throughout,
its thickness remains fixed, but since the deposited film behind is thinner,
the wave continually increases in volume.
This provides a mechanism for rapid wave growth in viscoplastic layers,
crucially reliant on the film being unyielded ahead of the wave.
The dynamics is potentially
distinct from that in periodic domains in which repeated traversals
inevitably adjusts the layer depth ahead of a wave
to that deposited behind (unless blow-up arises within a single transit).

Examples in which this phenomenon occurs can be identified in
figure \ref{fig:number_waves}(b,c), which shows results from further suites
of initial-value computations, taking \eqref{pertIC} as initial condition
with $A_b>0$. These examples are distinguished by blow-up occuring
after only one or two waves have formed for layers that would be
below the yield threshold if flat (to the left
of the vertical dotted lines). In fact, in the figure one sees
that such examples even persist to the right of the threshold, implying
that the mechanism continues to
operate when the entire fluid layer is above the threshold.
In order words, for a wave to rapidly amplify by entraining
upstream fluid whilst depositing less behind, the layer can also be
weakly yielded.
Note that,
for $\mS$ far beyond the yielding threshold, the dynamics becomes more like
that for a Newtonian layer, with multiple waves forming and the initial
bump having less impact. 


\section{Experiments}\label{sec:expts}

\subsection{Experimental setup}

\begin{figure*}[t!]
    \centering
    \includegraphics[width=0.46\linewidth]{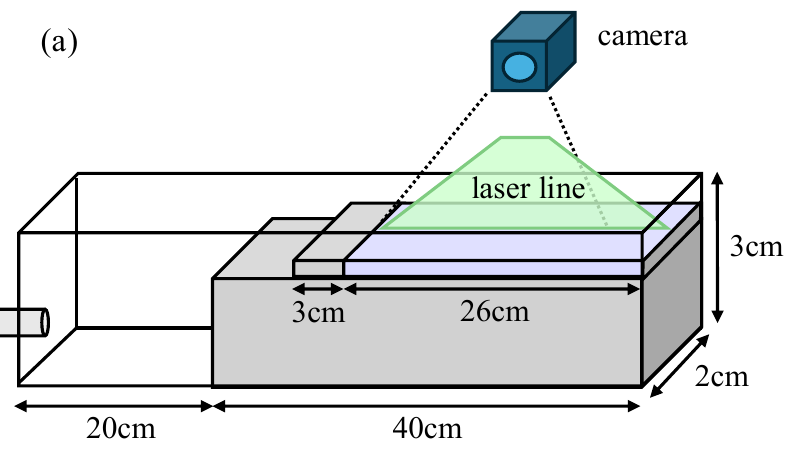}
    \vspace{12pt}
    \includegraphics[width=0.44\linewidth]{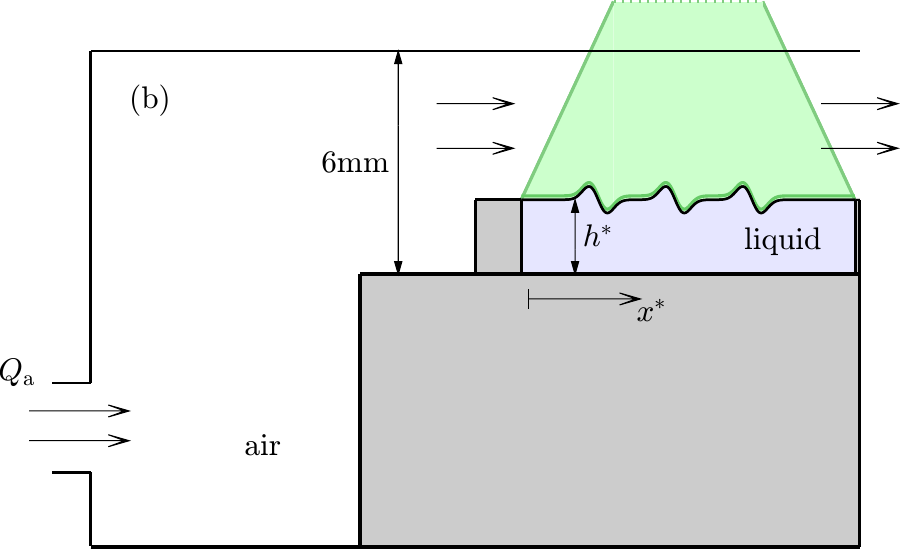}
    \includegraphics[width=.9\linewidth]{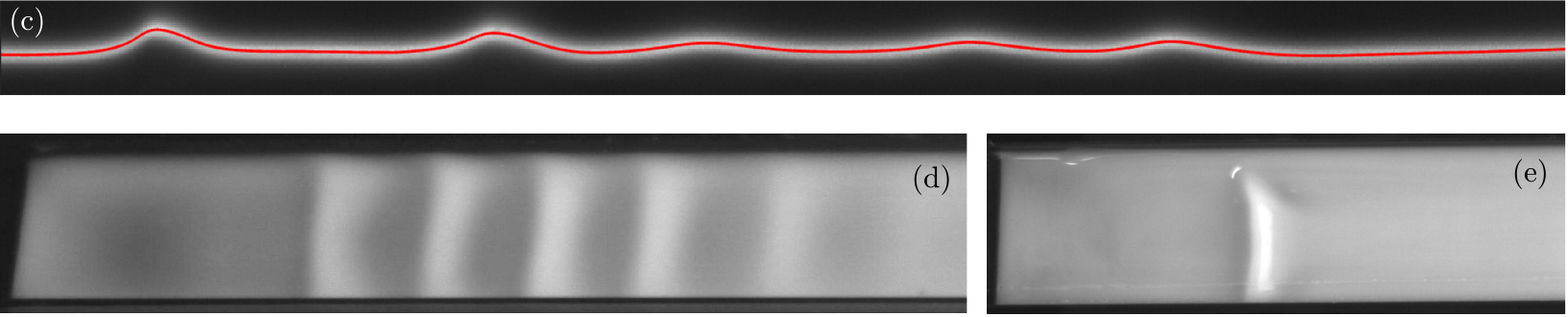}
    \caption{Sketches of the experimental setup
      showing (a) a persective of the entire tank and (b) a lengthwise
      cross-section. Air enters the tank with flux $Q_\mathrm{a}$ through
      an inlet on the right, and flows freely out of the other end.
      (c) Sample raw image of the surface of a glycerol layer, with the laser
      line illuminating the free surface along the centreline of the channel.
      The red line superposed is the fit to the centre of the laser line,
      which is used to determine the layer depth. Inclined images
      without the laser line but showing
      more of the liquid surface are displayed below for (d) a
      glycerol layer and (e) a gel layer. }
    \label{fig:intro}
\end{figure*}

The experiments were conducted in a rectangular acrylic tank of length $60$cm,
width $2$cm and depth $3$cm.
As sketched in figure \ref{fig:intro}(a,b), a PVC block was inserted
to divide the tank into a section on the left with length 
$20$cm and the full 3cm depth, and one to the right
with length $40$cm and 6mm depth.
To allow access, the roof of the tank was removable and held in place
using clamps during experiments.
The left end of the tank had an air inlet, connected to a compressed air line
fitted with a valve and flow meter to measure the flow rate, $Q_\mathrm{a}$.
The right end of the tank was open so that air could flow freely out. 

In preparation for each experiment, a liquid layer was emplaced above a
$26$cm-long section of the PVC block. Removable barriers were inserted on either
side to hold the layer in place.
In order to emplace layers with varying depth,
barriers of different heights were employed, ranging from $1$mm to $3$mm,
and scrapers were used to approximately level the liquid
surface. A laser line was shone through the roof of the tank onto the liquid
layer, illuminating a line down the centreline of the tank, parallel with the
side walls. The liquid surface was recorded using a
Jai Spark SP-5000M camera, at frame rates of up to $100$ frames per second
(fps). An approximately 18cm long section of the tank was recorded by the camera, which excluded the section closest to the right end of the tank (figure \ref{fig:intro}a). The camera was positioned at an angle of $39^\circ$ to the horizontal,
so that variations in the depth of the liquid layer could be captured. To
extract the midpoint of the laser line from an image, an 11-point parabolic
fit of the light intensity was applied at each pixel along the length of the
laser line. Figure \ref{fig:intro}(c) shows the laser
line captured during an experiment from a sample raw image, together with the line of fitted midpoints.
A Savitzky-Golay filter was applied to smooth this measurement
over a window of $18$ pixels; this level of smoothing was chosen to
reduce the level of noise in the data, which can inhibit identification
of small-amplitude waves, whilst adequately preserving the shape of
large-amplitude waves once formed. Measuring the shape of the interface
only along the tank centreline meant that the tranverse profiles of waves
were not generally captured. 
As illustrated in figure \ref{fig:intro}(d,e), waves typically appeared to
be relatively two-dimensional,
although some effects of the tank's side walls were visible.

\subsection{Materials}

We used two types of working liquid, one Newtonian and one non-Newtonian.
The Newtonian liquid was glycerol. The non-Newtonian liquid was a commercial
hair gel (Enliven Men Hair Gel Hold 2) diluted with water to various
concentrations. The key component of the gel is an aqueous solution of Carbopol, pH-neutralised by triethanolamine. Similar commercial gels have been used as model yield-stress fluids and characterised rheologically in previous studies
\cite{dinkgreve2016different,jalaal2019viscoplastic,taylor2024scraping}.
To dilute the gel, water was added to the desired concentration and
the suspension mixed for one hour using an electric mixer. A small quantity
of titanium dioxide was added to both working liquids to make them opaque. 

\begin{figure}[t!]
    \centering
    \includegraphics[width=\linewidth]{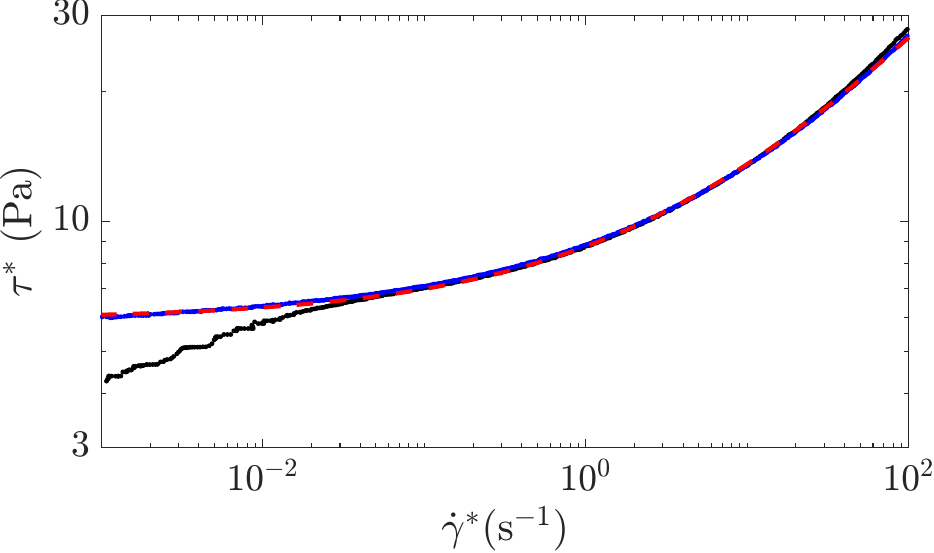}
    \caption{Shear-rate ramp of a gel/water sample, showing measurements on
      the upwards (black) and downwards (blue) ramps. The Herschel-Bulkley model fit to the downwards ramp data is shown as a dashed red line.
    }
    \label{fig:rheom}
\end{figure}

Controlled shear-rate tests were conducted  at $20^\circ$C in a Kinexus Pro+ rotational
rheometer (Malvern Instruments) using roughened parallel plates with a
$1$mm separation and $4$cm diameter.
Shear-rate ramps were conducted to measure flow curves for the 
different concentrations of gel mixture.
Following a similar protocol to that used by \cite{ribinskas2024scraping},
the shear rate was first increased from $10^{-4}\mathrm{s}^{-1}$ to
$10^{2}\mathrm{s}^{-1}$, then decreased over the same ranging.
The up and down ramps each lasted a total of $160$ seconds, with $500$
measurements taken. The Weissenberg-Rabinowitsch correction was used to
account for the non-uniform stress field generated in the parallel-plate
geometry \cite{mendes2014parallel}. The flow curve was then generated
from a rolling $10$-point median of the measured stress values.
A typical example is shown in figure \ref{fig:rheom}. The down ramp is
fitted using the Herschel-Bulkley model \cite{balmforth_yielding_2014}
over the range $10^{-3}\mathrm{s}^{-1}$ and $10^{2}\mathrm{s}^{-1}$.
There were generally good fits with the Herschel-Bulkley model and minimal
difference between the up and down ramps suggesting little thixotropy. More
extensive rheological tests have been carried out on the same brand of
commercial gel by Taylor-West \& Hogg \cite{taylor2024scraping}, including
oscillatory measurements to characterise elastic properties, although the
gel was not diluted in that study. 

There was some variability in the flow curves of the gel mixtures depending
on the amount of time between preparing the fluid (diluting with water and
mixing) and conducting the rheometry tests.
Four different concentrations of gel were used;
for each, we conducted a rheological test the day before experiments,
and then a second test was conducted whilst the experiments were being
conducted with that fluid or soon afterwards. In one case, the second test was conducted the day following experiments being conducted with that sample. Table \ref{tab:expts_rheo} shows
the Herschel-Bulkley parameter values fitted in each of these tests for
all four concentrations. Significant variations in $\tau_{\mathrm{y}}$
are observed between the two tests, implying some uncertainly
in the yield stress of the fluid in the experiments.
Repeating rheological tests in quick succession, rather than after waiting
a day, yielded reproducible results, suggesting the variations were not
simply due to liquid inhomogeneity or inadequate mixing. 
Given the relatively small samples of fluid used, 
we assume that the variations are more likely due to evaporation.
Therefore, we assume that the yield stress of the fluid used in an
experiment lies approximately within the range of values given
in table \ref{tab:expts_rheo}.


We also measured the viscosity of the glycerol: before being used
in an experiment, the viscosity was $1.1$ Pas. We also tested glycerol samples
after they had been used in a number of experiments, finding that the
viscosity had typically dropped, with the lowest viscosity recorded being
$0.5$ Pas. This may be due to contamination of the glycerol with water during
the experiments, as glycerol is hygroscopic and glycerol/water viscosity is
strongly dependent on the concentration of water \cite{ault2023viscosity}.

In calculations below, we assume the surface tension of glycerol to be
$63\mathrm{mNm}^{-1}$, and its density to be $1260\mathrm{kgm}^{-3}$.
We take air to have viscosity,
$\nu_\mathrm{a}=1.5\times10^{-5}\mathrm{m}^2\mathrm{s}^{-1}$,
and density, $1.2\mathrm{kgm}^{-3}$.
The channel depth was $H=6$mm, and the channel width was $W=20$mm. We assume that the 2D air flux, $Q$, used in the long-wave model, is equivalent to $Q_\mathrm{a}/W$, where $Q_\mathrm{a}$ is the volume flow rate of air into the tank. The friction coefficient, $\mf$, is relatively poorly
constrained, but we may approximate it crudely from Moody plots
\cite[e.g., ][]{schlichting2016boundary}, as was done by Basser et al.
\cite{basser_1989_cough}, which suggests a value of roughly $\mf=0.005$ for
the typical Reynolds numbers used in our experiments. We also conducted an experiment
to provide further evidence for this estimated value of $\mf$
in the current setup: we scattered tracer particles on the surface of a
glycerol layer with depth of approximately $1$mm and subjected it to air
flow with a rate of $Q_\mathrm{a}\approx0.52\mathrm{Ls}^{-1}$, which is
low enough that surface waves did not form. Averaging the distance travelled
by several tracer particles over $40$ seconds leads to an estimate of the
surface velocity. A
comparison with the prediction of the long-wave model for a uniform
Newtonian layer, then suggested an approximate value of $\mf\approx0.006$
when the glycerol viscosity was assumed to be $\eta=1.1$Pas, although there was some variability in the speed of tracer particles at different locations on the layer.
Overall, we cannot expect the single-parameter air-flow model to accurately
capture the dynamics of the air layer regardless of the value of $\mf$.
However, given an estimate for $\mf$, we may make some 
comparisons between theory and experiments of a more quantitative
flavour.

\begin{table}
    \centering
    \begin{tabular}{cc|ccc}
        Sample & Test no. & $\tau_{\mathrm{y}} (\mathrm{Pa})$ & $n$ & $K (\mathrm{Pa} \cdot \mathrm{s}^n)$ \\
        1 & 1 & 0.6 & 0.47 & 0.8 \\
        1 & 2 & 1.0 & 0.52 & 1.0 \\
        2 & 1 & 2.4 & 0.45 & 1.6 \\
        2 & 2 & 3.0 & 0.47 & 1.8 \\
        3 & 1 & 1.8 & 0.44 & 1.3 \\
        3 & 2 & 4.2 & 0.50 & 2.3 \\
        4 & 1 & 4.8 & 0.44 & 2.3 \\
        4 & 2 & 6.0 & 0.49 & 2.7 \\
    \end{tabular}
    \caption{Measured properties of the four different gel/water mixtures used in experiments, after fitting to the Herschecl-Bulkley model. The fitted quantities are the yield stress, $\tau_\mathrm{y}$, power-law index, $n$, and consistency, $K$ \cite{balmforth_yielding_2014}. For each sample, results from two rheological tests are given, with the fluid having been left in a beaker for 1 day (samples 1, 2, 4) or 2 days (sample 3) between Test 1 and Test 2. }
    \label{tab:expts_rheo}
\end{table}

\subsection{Air flow protocols}

\begin{figure*}[ht!]
    \centering
    \vspace{6pt}
    \begin{subfigure}{0.55\textwidth}
        \centering
        \includegraphics[width = \textwidth]{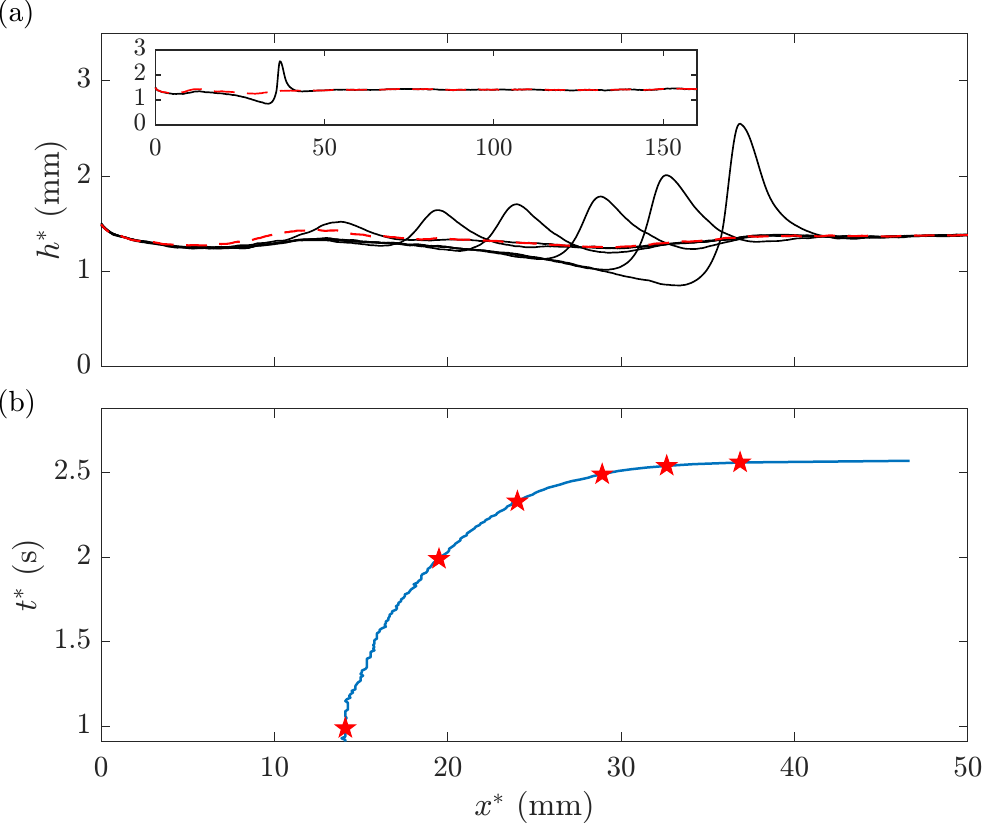}
    \end{subfigure}%
    \ \ \
    \begin{subfigure}{0.3\textwidth}
        \centering
        \includegraphics[width = \textwidth]{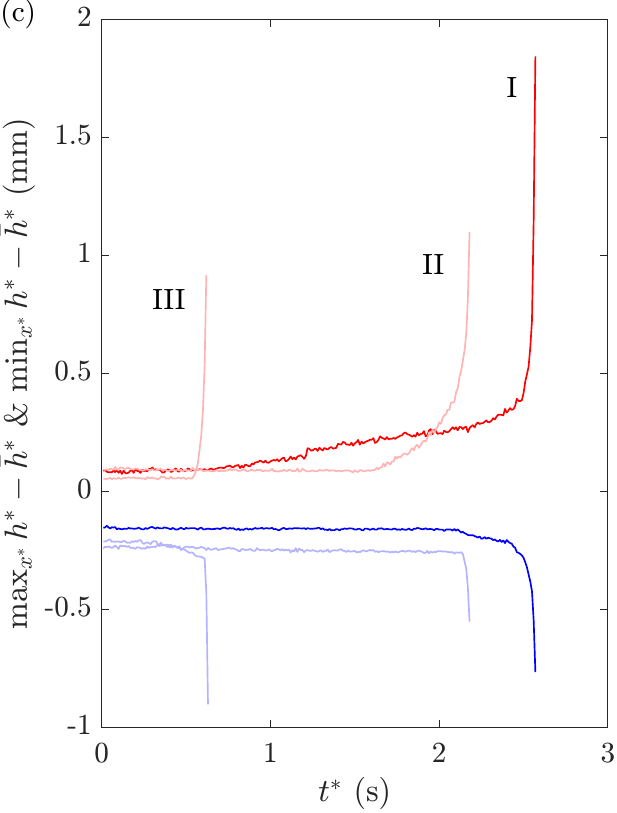}
    \end{subfigure}
    \caption{(a) Snapshots of the liquid interface from a gel experiment in
      which the air flow rate was turned up suddenly from zero to
      $Q_\mathrm{a}\approx1.2$L/s. The red dashed line shows the initial interface.
      The inset shows the first and last snapshots, zoomed out to show
      more of the downstream interface. (b) Spatial location of the
      maximum layer height. The stars indicate the snapshots plotted in (a).
      (c) Time series of the maximum (red) and minimum (blue) layer height
      relative to the initial mean height, $h_\mathrm{av}\approx1.4$mm,
      (denoted I). Also included are data from two
      repetitions of the same experiment (labelled II and III).
      In all experiments, liquid made contact with the tank roof immediately
      after the last time plotted in (c).}
    \label{fig:VPeg}
\end{figure*}

We conducted experiments using two protocols for setting
or adjusting the air flow rate. The first protocol (protocol I) was to suddenly increase
the air flow rate from zero up to the desired rate. This was achieved by
first increasing the pressure at the air compressor while the tubing was disconnected from the tank until the desired air flow rate was achieved.
The valve was then closed and the tubing reconnected to the tank. Finally,
the valve was opened to allow air to flow through the tank, and the interface
shape was recorded, with the air flow continuing for five seconds or
until the liquid made contact with the roof of the tank. This protocol was
used for the glycerol experiments and several experiments with
gel (figures \ref{fig:expts_NQ7}-\ref{fig:VPeg}). The valve was operated
manually, and fully opening it took up to one second. We estimate the typical errors in setting the final flow rate to be up to $0.02\mathrm{Ls}^{-1}$. 

\begin{figure}[t!]
    \centering
    \includegraphics[width=\linewidth]{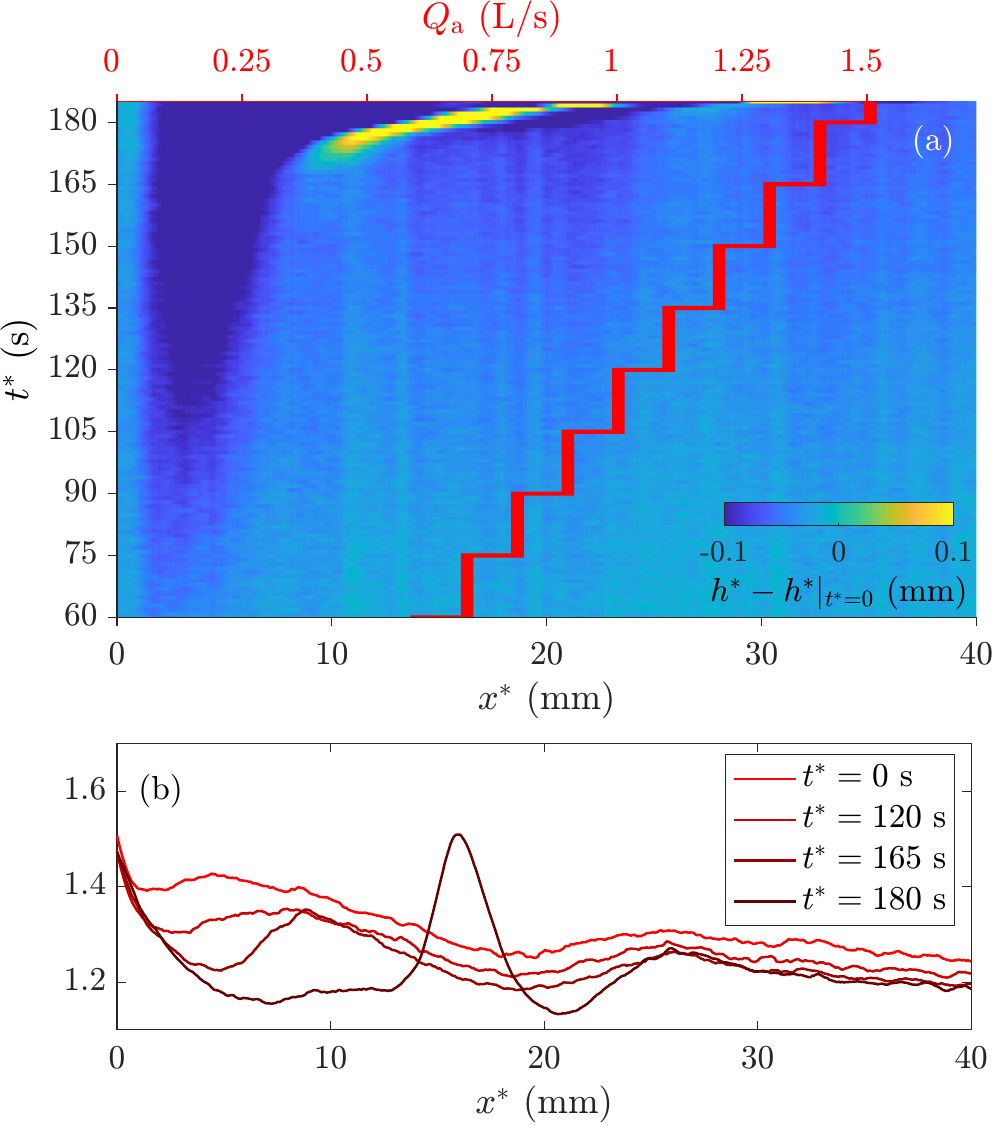}
    \caption{An experiment with a gel layer (sample 2) in which the air flow rate was increased in increments every $15$s. (a) Layer depth relative to initial depth. Dark blue and yellow regions indicate where the depth has changed by at least $0.1$mm. Red steps indicate the air flow rate. (b) Four snapshots of the interface shape. Images show only a section of the layer close to the tank inlet; further downstream there is no appreciable deviation in the layer depth.
    }
    \label{fig:creep}
\end{figure}

The second protocol (protocol II) was used only for gel experiments
(figures \ref{fig:creep} and \ref{fig:gelcrit}), and involved 
increasing the air flow rate in a more gradual manner, in fixed
steps of either $0.09\mathrm{Ls}^{-1}$ or $0.13\mathrm{Ls}^{-1}$,
taken every five or fifteen seconds. Each time,
the flow rate was increased until liquid made contact with the roof of the
tank. A similar protocol was by Basser et al. \cite{basser_1989_cough}.
The second protocol was not used for glycerol
because that fluid could be swept through the tank at low flow rates,
implying that layer depths changed significantly before
flow rates became sufficient to generate surface waves. 

\subsection{ Newtonian liquid layers}

\begin{figure}[t!]
    \centering
    \includegraphics[width=\linewidth]{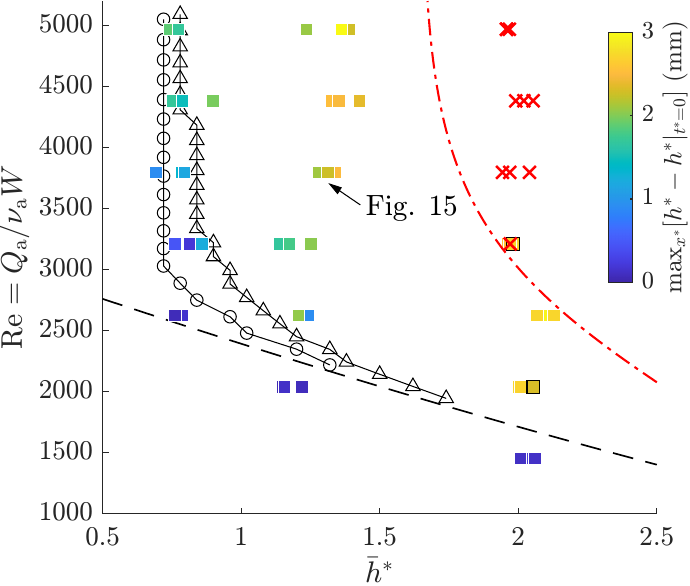}
    \caption{Maximum layer height relative to initial height recorded for
      each experiment with glycerol (coloured symbols). Red crosses indicate
      experiments where a blow-out of liquid onto the roof of the tank occurred. Experiments where a few small droplets of liquid hit the roof of the tank, but a significant blow-out did not occur, are marked as coloured symbols enclosed by black squares. 
      The red dot-dashed line is the rough boundary between
      experiments in which blow-out did or did not occur.
      The dashed black line is the linear instability
      threshold predicted by \eqref{LSA_omr}. Solid black lines
      show the critical layer depths for which blow-up occurs
      in periodic initial-value computations with the long-wave model,
      using $\mf=0.005$ (circles) and $\mf=0.01$ (triangles).
    }
    \label{fig:glyc_maxamp}
\end{figure}

Figure \ref{fig:glyc_maxamp} displays a regime diagram of the dynamics
observed for layers of glycerol with different depth and air flow rate.
When the air flow rate was sufficiently low, the maximum layer height did not
appreciably increase from the initial film thickness, suggesting that there
was no surface instability;
these experiments are shown by blue squares
in figure \ref{fig:glyc_maxamp}. For somewhat higher flow rate,
surface waves were generated, with maximum recorded amplitudes
indicated by the coloured squares in figure \ref{fig:glyc_maxamp},
without any waves hitting the roof of the tank, although in a small number of cases (indicated by black squares in the figure), a few small droplets of fluid were generated as a wave was propagating which did end up on the tank roof. 
Finally, when the layer thickness and flow rate were high enough, waves formed that grew until they made contact with the tank roof,
{\it i.e.} ``blow-out'' occurred
(these cases are shown by red crosses in the figure). 

\begin{figure*}[t!]
    \centering
    \includegraphics[width=0.8\linewidth]{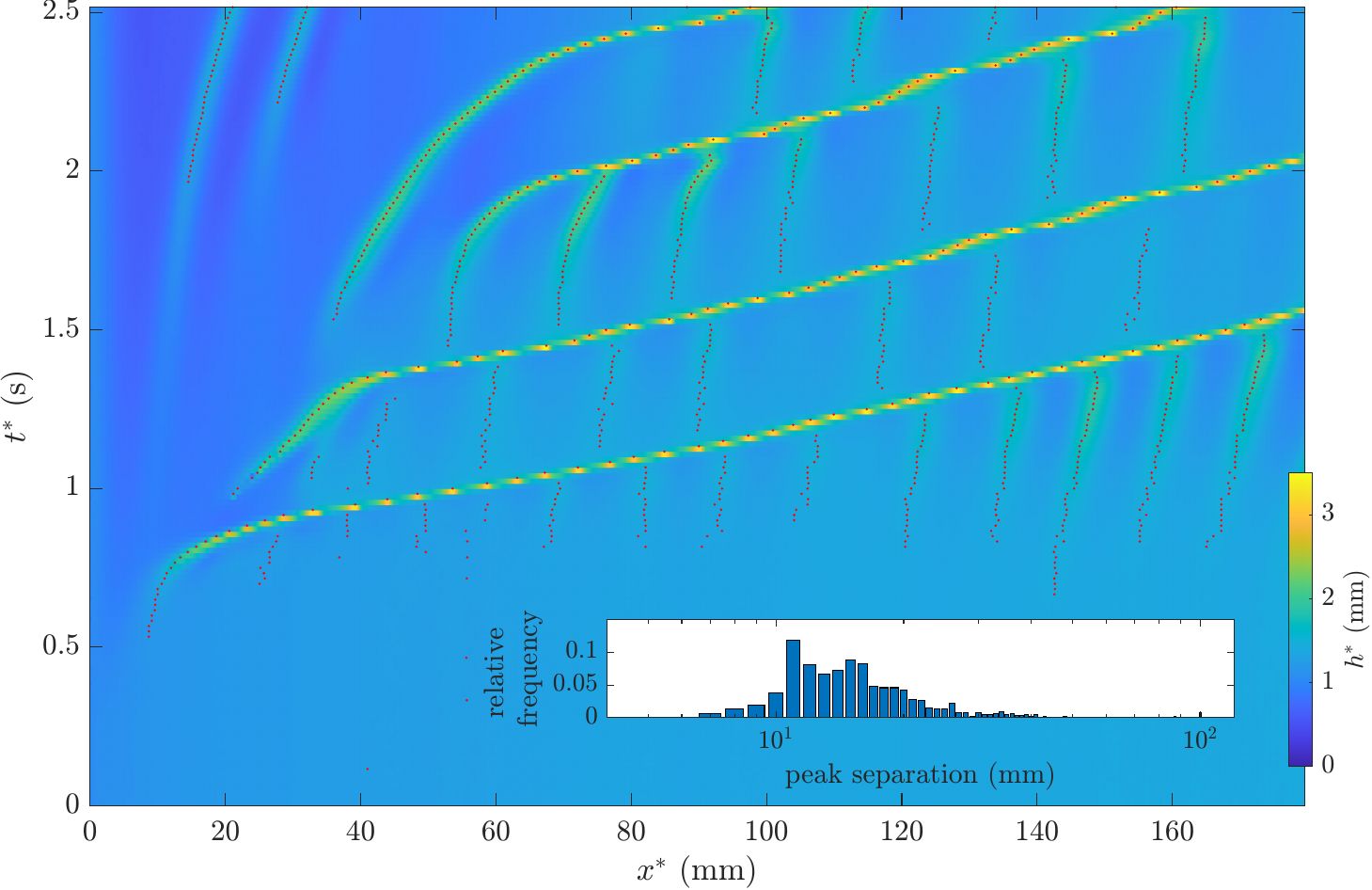}
    \caption{(a) Glycerol film height in an experiment with an air flow rate
      of $Q_\mathrm{a}\approx1.14$ L/s and an initial mean layer depth of $\bar{h}^*\approx1.3$mm.
      Red dots indicate every point in space and time where a peak is
      detected (see main text for details).
      The inset shows a histograms of all the instantaneous
      peak separations thereby identified.
    }
    \label{fig:expts_NQ7}
\end{figure*}

The critical flow
rate required for the growth of surface waves depends on the mean layer
thickness and can be predicted relatively well by the linear stability
analysis of the long-wave model in \eqref{LSA_omr}, provided that gravity is
taken into account (see figure \ref{fig:glyc_maxamp}).
The rough border between the regimes in which waves saturate or
blow-out occurs is drawn as a dot-dashed line
in figure \ref{fig:glyc_maxamp}. This second threshold also depends
on the layer thickness and is only poorly predicted
by the long-wave model, if it is assumed that blow-out is
equivalent to finite-time blow-up in the latter
({\it cf.} \S\ref{sec:nonlinear_periodic} and figure \ref{fig:carp}):
with $\mf=0.005$, the model predicts the
threshold indicated by open circles in figure \ref{fig:glyc_maxamp}.
Even if the friction factor is increased arbitrarily to
$\mf=0.01$, the predicted threshold (open squares) still remains
well to the left of that observed experimentally.
Thus, the Newtonian long-wave model predicts blow-up in cases where
experiments exhibit large-amplitude waves that propagating stably until they
reach the end of the tank or recording stopped.
This weakness of the model
is perhaps expected given previous discussion of similar models for two-layer
or inclined plane flow of Newtonian fluids \cite{dietze2013wavy}.
Nevertheless, the existence of the three regimes in figure \ref{fig:glyc_maxamp}
are at least qualitatively predicted by the long-wave model,
suggesting that it remains a useful complement to experiments
even if it is not a quantitative tool. 

Figure \ref{fig:expts_NQ7} shows further details of the dynamics in an
experiment in which surface waves were generated without blow-out.
At early times, a train of waves is generated with
relatively even spacing between the wave peaks. One or a small number of
those waves then grows notably faster than the rest.
In figure \ref{fig:expts_NQ7}, the largest waves are mostly seeded from
near the solid barrier at $x^*=0$, and the largest wave travels fastest,
coalescing with smaller waves in the wave train ahead of it. As time
progresses, the film thickness noticeably
drops near $x^*=0$, as fluid becomes
swept away from the barrier towards the right end of the tank. 
Computations with the model using a zero-flux boundary condition at $x=0$
qualitatively reproduce this thinning trough, as well as the wave train
that emerges nearby (see figure \ref{fig:Nlong}). However, whilst
the model does in some cases predict wave coalescence,
the interactions appear somewhat different, with
the model often predicting blow-up immediately after a coalescence
(e.g., figure \ref{fig:Nlong}c). 

To quantify in more detail the waves
in experiments like that shown in figure \ref{fig:expts_NQ7},
we identify a wave peak as any point in space, $x^*=x_i$, at which the
film height, $h^*(x_i,t^*)$, is a local maximum, to within $5$mm in either
direction, and provided $h^*(x_i,t^*)-h^*(x_i,0)>A_0$, for a
some minimum depth $A_0$.
Practically, we take $A_0=0.02$mm, which is small enough that early-time
waves with relatively small amplitudes can still be detected, but sufficently
large that noise in the depth measurements does not generate a
significant number of spurious identifications. The peaks located
for the experiment in figure \ref{fig:expts_NQ7} are shown by red dots,
and a histogram of all the peak separations thereby identified
is included as an inset. The broad, but fairly well-defined maximum
of that distribution of separations between $11$mm and $16$mm
reflects the typical spacing between waves at early times, and
is suggestive of a preferred wavelength to the surface-wave instability.
A few of the measured peak separations are significantly larger than $16$mm,
and likely reflect the fact that peaks do not all emerge at the same
time, with some separated by gaps that become filled by new peaks
at later times. 

\begin{figure*}[t!]
    \centering
    \vspace{6pt}
    \begin{subfigure}{0.44\textwidth}
        \centering
        \includegraphics[width = \textwidth]{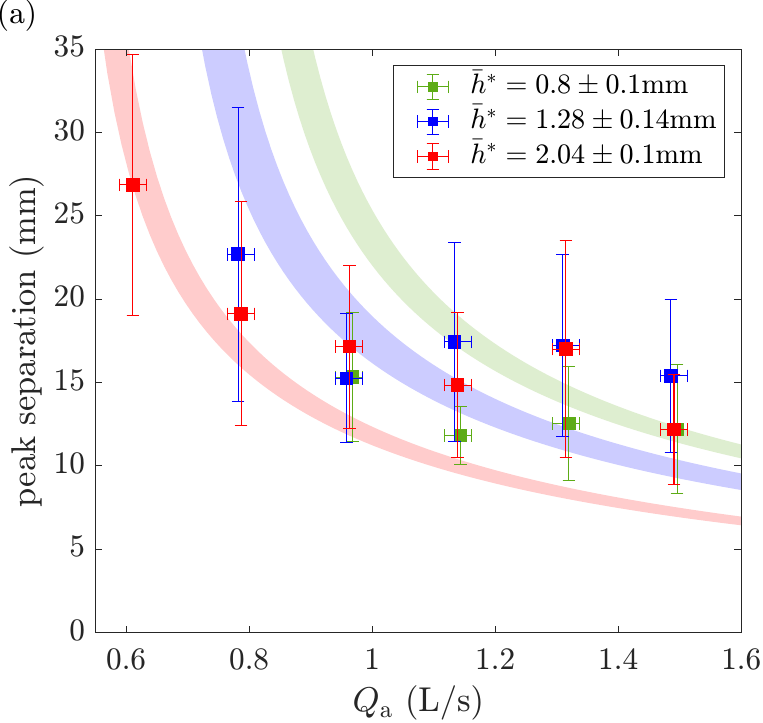}
    \end{subfigure}%
    \begin{subfigure}{0.44\textwidth}
        \centering
        \includegraphics[width = \textwidth]{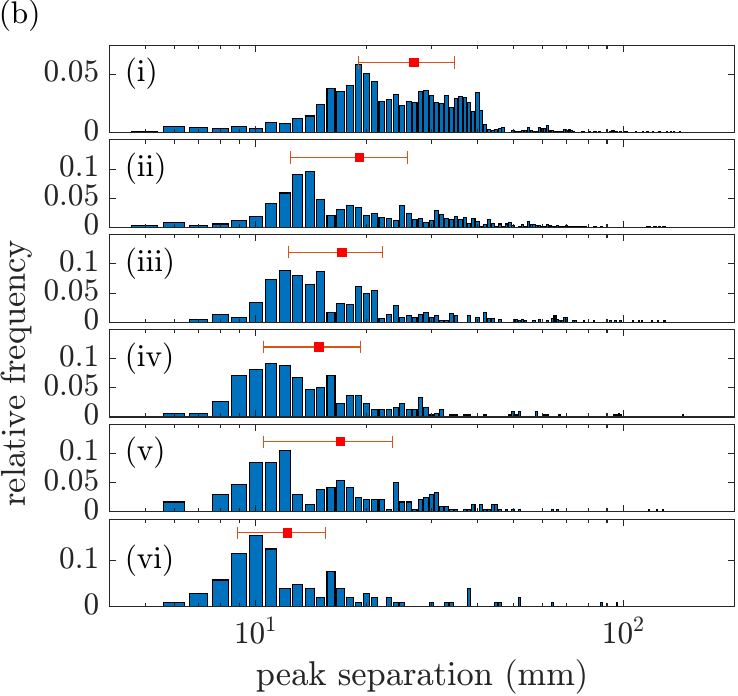}
    \end{subfigure}
    \caption{(a) Median peak separations for varying air flow rate, $Q_\mathrm{a}$, and initial mean layer thickness, $\bar{h}^*$. Vertical error bars correspond to the median absolute deviation (MAD) in the peak separation data. Horizontal error bars reflect the experimental error in achieving the desired air flow rate. Bands of solid colour indicate the predicted wavelength from linear theory \eqref{LSA_omr}, with the width of the bands reflecting the ranges of values of $\bar{h}^*$ for each suite of experiments. (b) Distribution of measured peak separations for all experiments with $\bar{h}^*\approx2.0$mm. Air flow rate increases from (i) $Q_\mathrm{a}\approx0.61$L/s to (vi) $Q_\mathrm{a}\approx1.5$L/s. The separations are extracted from all experiments conducted at that given flow rate. In (i)-(vi), the medians and
      their MAD error bars, are plotted in above the histograms,
      and correspond to the red data points in (a).
      }
    \label{fig:glyc_wl}
\end{figure*}


Peak separations from all our experiments are assembled
in figure \ref{fig:glyc_wl}a. Here, we plot the median of
the separation distributions, as illustrated in the inset
of figure \ref{fig:expts_NQ7} and for a suite
of experiments with approximately fixed initial layer thickness in
figure \ref{fig:glyc_wl}b. Assuming that the median
does indeed correspond
to the preferred wavelength of linear surface-wave instability,
figure \ref{fig:glyc_wl}a compares the experimental data with the
predictions of the long-wave model from \eqref{LSAkm}.
There is order-of-magnitude
agreement between the theory and experiments, despite
a significant spread in the measurements ({\it cf.}
figure \ref{fig:glyc_wl}b). Similarly,
the instantaneous speeds of low-amplitude peaks extracted from plots
like figure \ref{fig:expts_NQ7} share the order-of-magnitude
of the speed scale
$\rho_{\mathrm{a}}\mf Q_\mathrm{a}^2 / (\eta HW^2) \approx 5$mm/s
identified in the model. 
However, the observed wavelength seems less sensitive
to mean layer depth, $\bar{h}^*$, than expected theoretically
(figure \ref{fig:glyc_wl}a).





\subsection{Yield-stress liquid layers}



Snapshots of the interface from a sample experiment with gel
and ramping up the air flow using protocol I
are shown in figure \ref{fig:VPeg}(a). 
In contrast to typical Newtonian experiments,
only a single isolated surface wave forms, with the remainder
of the layer remaining largely in place.
That wave grows and propagates at an accelerating rate. Unlike for
waves in Newtonian films with similar depths and air flow rates
(figure \ref{fig:expts_NQ7}), the wave amplitude does not saturate,
but blow-out is triggered (see figure \ref{fig:VPeg}b,c). 
 
For the example 
in figure \ref{fig:VPeg}(a), the wave appears to be seeded initially
from a small bump in the free surface located near $x^*\approx12$mm.
This imperfection in the initial film is apparently sufficient for the
fluid to yield locally around this bump and create the wave
before any significant yielding elsewhere in the layer. Moreover,
as the wave grows and propagates, the film deposited behind the wave
becomes gradually thinner, whereas the layer ahead remains close
to its initial depth. Thus, as the wave propagates, the volume of the wave
grows continually, triggering explosive nonlinear growth
and blow-out as in our initial-value computations
in long domains (\S\ref{sec:longY}).

As illustrated in figure \ref{fig:VPeg}(c), which shows three repetitions
of the same experiment, this dynamics is typical of a gel layer
with a fast ramp up of the air flow rate.
Evidently, in the initial preparation of the layer, for which a scraper
is used to flatten the surface, small depth perturbations are unavoidable.
The resulting imperfections act like the localised bumps introduced
into our initial-value computations, and are visible
in figure \ref{fig:VPeg}(c) as the
differences between the maximum and minimum layer depths for $t^*\to0$.


In experiments in which the air flow rate was gradually increased over a
longer time (protocol II), we still found the same qualitative wave dynamics:
a single wave formed when the flow rate reached some threshold; the wave
then grew explosively and led to a blow-out event.
However, in some of these experiments, there was a measurable, gradual
decrease in the layer depth close to the boundary at $x^*=0$ at flow rates
well below the blow-out event.
The example in figure \ref{fig:creep} illustrates this
apparently sub-yielding deformation: the figure displays
the evolution of the layer depth, along with a time series of the
flow rate. As the flow rate is ramped up, a trough forms near $x^*=0$
long before any measurable wave generation. Somewhat like
in the Newtonian version of the problem, this trough assists in providing
a localised perturbation to the free surface. The wave
that blows out the channel forms on the
right-hand shoulder of the trough near $x^*=10$mm.

Thus, the initial preparation of the layer was not the only factor
in determining where the layer locally yielded to nucleate waves.
In most experiments,
we found that the wave formed near to the boundary at $x^*=0$,
suggesting either that an inhomogeneity was present there from the outset,
or that a trough formed near the boundary via sub-yield deformation,
as in figure \ref{fig:creep}. We also cannot discount the possibility
that modifications to the turbulent air flow through the tank, due to
some geometrical feature of the channel or the change in surface
properties across the barrier,
systematically generates a higher stress on the liquid near $x^*=0$.


The physical origin of the sub-yielding deformation
evident in figure \ref{fig:creep} is not completely clear.
It is certainly common for real fluids to display much more
complicated material behaviour in the vicinity of the yield stress
than is captured by the Bingham model,
and even to possess no true yield threshold
\cite{balmforth_yielding_2014,bonn2017yield}.
The flow curves in our rheometry do not rule out deformation
below the fitted yield stress, and may suggest a mild
viscous response (see also \cite{taylor2024scraping}).
Curiously, the creeping motion exposed by figure \ref{fig:creep}
does not noticeably adjust at each ramping up of the flow rate,
which constrains any viscoelastic response.

\begin{figure}[t!]
    \centering
    \includegraphics[width=\linewidth]{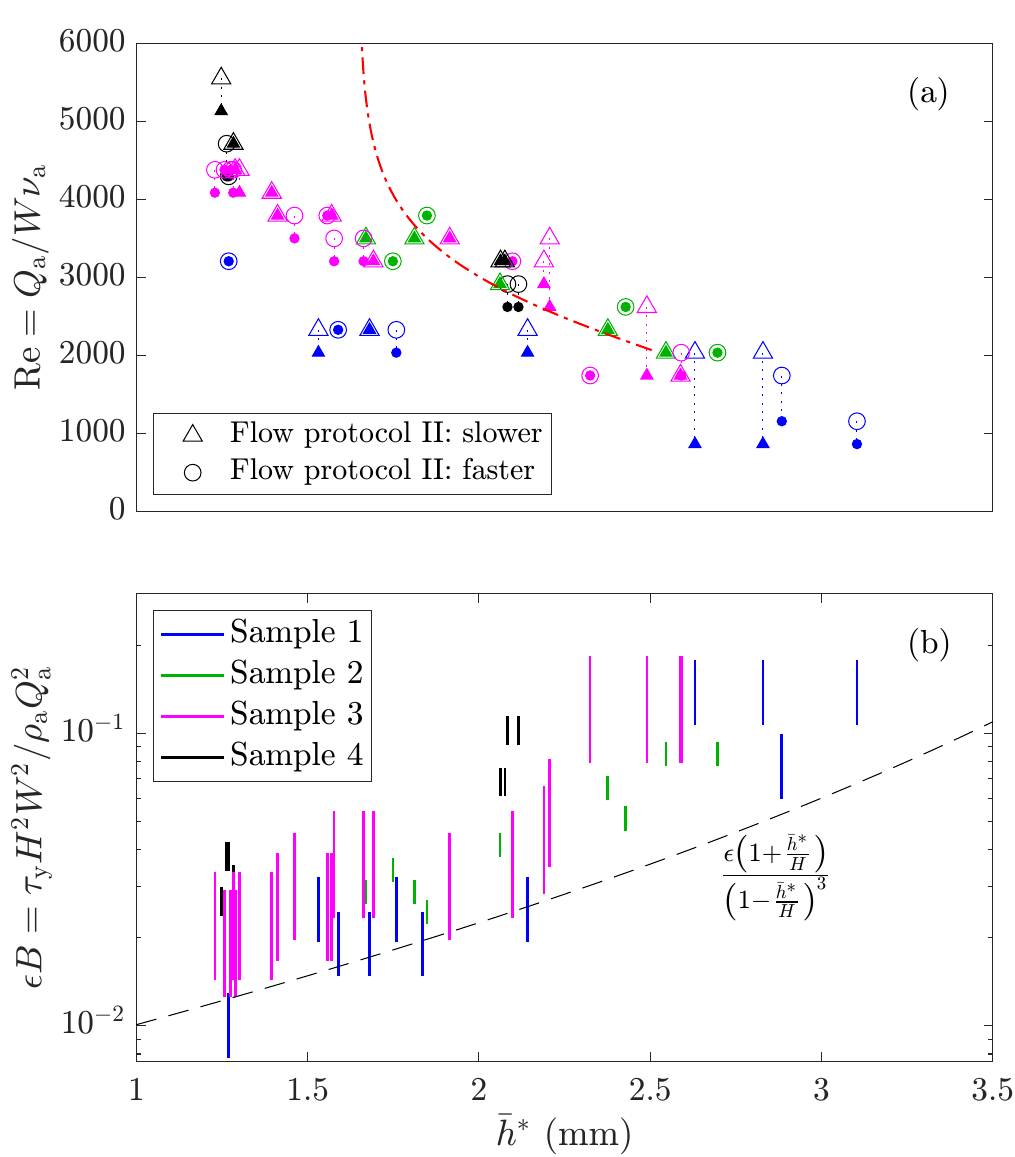}
    \caption{(a) Critical Reynolds number
      Re$=Q_\mathrm{a}/(W\nu_\mathrm{a})$
      to induce motion in experiments
      with gel layers.
      The solid symbols record the $\mathrm{Re}$ at
      which $h^*-h^*|_{t^*=0}$ first exceeded $0.1$mm somewhere in the layer;
      the open symbols indicate the flow rate at which liquid hit the roof
      of the channel. Each pair of data points is connected by a dotted line.
      Triangles indicate an experiment in which the air flow rate was
      increased every fifteen seconds; in the remaining experiments
      (circles), the flow rate was increased three times faster.
      Colours correspond to different gel concentrations,
      as indicated.
      The
      red dot-dashed line reproduces the Newtonian regime boundary
      from figure \ref{fig:glyc_maxamp}.
      (b) Critical Bingham number, $B$, corresponding to the air
      flow rate for which max$(h^*-h^*|_{t=0})>0.1$mm. The vertical length
      of each line represents the range of measured values for the
      yield stresses (table \ref{tab:expts_rheo}). The dashed line
      shows the yielding threshold, $B=(1+\bar{h})/(1-\bar{h})^3$,
      using $\mf=0.005$. 
    }
    \label{fig:gelcrit}
\end{figure}

Given that perturbations to the initial layer depth or sub-yield deformation
both seem able to trigger localised yielding, it seems reasonable to assume that
the theoretical yielding threshold for an exactly uniform film provides
an upper bound on the air flow rate required for wave formation.
In figure \ref{fig:gelcrit}, we compare that threshold with results from
all of the gel experiments conducted using protocol II. The raw data,
shown in panel (a), presents flow rates translated to air Reynolds number
against layer depth. For each experiment, two data points are shown: the
first records the flow rate at which significant motion was first
detectable in the layer, defined as a depth increase exceeding $0.1$mm.
The second point corresponds to the flow rate at which blow-out occurred.
For most of the experiments, the pair of data points coincide, reinforcing
our earlier conclusion that significant yielding immediately
leads to blow-out. The raw data for the first data set is translated
to a critical Bingham number in panel (b), and shows how
the theoretical threshold does indeed largely bound the observations
when we take $\mf=0.005$, even taking into account the uncertainty
in the yield stresses of the fluid. 
Note that relatively thin layers of yield-stress fluid still exhibit blow-out, even for film thicknesses where only stable waves were observed for glycerol layers (figure \ref{fig:gelcrit}a). 
In other words, catastrophic blow-out events occur even for relatively thin viscoplastic films, in complete contrast to Newtonian ones.


\section{Discussion}\label{sec:discussion}

In this paper, we have presented a theoretical model
and experiments exploring the air-driven surface-wave instability
of a layer of viscoplastic fluid. A main goal was to
interrogate the effect of a yield stress. In an earlier paper,
Basser {\it et al.} \cite{basser_1989_cough} presented
some exploratory experiments and
physical arguments to suggest that a yield
stress could promote an instability that could lead to
a violent ``blow-out'' of the viscoplastic lining of a duct.
Here, we have investigated this possibility in far more detail,
confirming that the yield can indeed have this effect. The
actual mechanism for blow-out is a little different from
the image presented by Basser {\it et al.}, however, with
the phenomenon driven by localised surface
waves consuming the upstream viscoplastic film in a runaway growth,
rather than an avalanche-like build up reliant on slip (that said,
slip may nevertheless have played an important role in
Basser {\it et al.}'s experiments and might be relevant in
mucus clearance in the lung).

Our model combines lubrication theory for the flow of the liquid film with 
a St-Venant-type model with a Ch{\'e}zy drag law for
the air flow. While this model may oversimplify turbulent air-flow dynamics
and the interaction with the liquid film, it captures key effects required
for surface-wave generation: the Bernoulli forcing from air inertia,
turbulent drag and surface tension. The model provides a relatively simple theoretical framework to investigate the impact of viscoplastic rheology on surface-wave dynamics. 
In particular, we have demonstrated how small perturbations in layer
depth can induce localised yielding in the liquid film from which
isolated waves can become nucleate. By consuming the
unyielded fluid in the film ahead, but leaving
a much thinner deposit behind, the waves
grow explosively, triggering a finite-time blow-up in the model.
Although the model cannot reliably capture such blow-up dynamics,
it is tempting to associate this singular behaviour
with the violent blow-out seen in experiments.

In our experiments,
we generally observed multiple surface waves forming
on Newtonian films. Those waves propagated to the end of the tank, or
accelerated to hit
the tank roof and caused a blow-out event when the film thickness and
air flow rate were sufficiently high. The long-wave model provides
reasonable estimates of the critical air-flow rate required
to initiate surface instability
(figure \ref{fig:glyc_maxamp}), as well as their wavelengths
(figure \ref{fig:glyc_wl}). However, the critical film thickness
for blow-out in the experiments is rather different to that needed
for finite-time blow-up in the model (figure \ref{fig:glyc_maxamp}).
Moreover, nonlinear wave interactions proceeded differently in the model
({\it cf.} figure \ref{fig:Nlong}c and figure \ref{fig:expts_NQ7}).
These discrepancies probably reflect limitations in the treatment of
air flow in the model. However,
it is also possible that finite-time blow-up does
not correspond to experimental blow-out, rather only a failure of the
long-wave framework, as has been discussed for related shallow-flow models \cite{pumir1983solitary,ruyer2002further,dietze2013wavy}.

In experiments with yield-stress fluid, we typically observed dramatic
growth of isolated waves. As in the model, such waves grew explosively from
locally yielded surface perturbations (consuming the
film ahead and leaving a shallower deposit behind), and
consistently generated blow-out events ({\it cf.} figure \ref{fig:gelcrit}a
and figure \ref{fig:VPeg}). However, we also
found evidence for surface deformation at air flow rates
well below those needed to trigger significant yielding.
The origin of this sub-yield deformation is not clear, but viscous
creep or elastic deformation below the yield stress are both possible
in Carbopol-based gels \cite{bonn2017yield}.

Our results highlight
the key role played by the liquid yield stress in generating dramatic wave
growth and precipitating significant clearance events (blow-out)
of the viscoplastic lining of a duct. 
It is also clear that, if the liquid yield stress is too high, motion
is entirely suppressed. In the context of airway clearance,
intermediate mucus yield stresses, relative to wind stress, are then
associated with the most efficient rates of transport.
This observation may have implications for the effectiveness of treatments
for obstructive lung diseases, such as airway clearance techniques that
involve forced expiration \cite{belli2021airway} and inhaled 
drugs that alter the mucus yield stress \cite{patarin_rheological_2020}.
However, many key physiological features of airways must be incorporated
if a physiologically complete model of cough is to be developed, including
airway wall compliance, non-uniform wall geometries, and shorter bursts of
air flow mimicking a typical cough. 

As our focus in this study has been on examining yield-stress effects,
we have used the Bingham constitutive law for the model. However, real
airway mucus is also viscoelastic, and most likely thixotropic
\cite{lai_micro-_2009,quraishi_rheology_1998}. Previous experimental models of cough have suggested that viscoelasticity and thixotoropy can alter liquid transport
rates \cite{zahm1991role,king1987role}. Extending our model to incorporate an elastoviscoplastic or thixotropic constitutive law may elucidate how these rheological properties may impact the dynamics of air-driven mucus clearance by coughing. 

Experimentally, modifying the setup for faster air flow would allow
the use of liquids with higher yield stresses.
The highly diluted gel samples we used here appeared to be prone to
evaporation, an effect that would be alleviated with a higher yield stress.
Similarly, using alternative working liquids that more closely mimic
mucus would also provide further physiologically relevant insight.

\section*{Acknowledgements}
JDS has been supported by a Leverhulme Trust Study Abroad Studentship and by the EPSRC Network Plus `BIOREME' (EP/W000490/1). We thank D. M. Martinez for helpful discussions. 

\appendix

\section{Regularised model flux}\label{app:reg}

From \eqref{constit_regGNF}, 
\begin{equation}
    \tau(\dot\gamma+\delta) = \dot\gamma\left(\dot\gamma+\delta+B\right),\label{GNFconstquadratic1}
\end{equation}
where, to leading order, $\dot\gamma = |\dot\gamma_{xy}|=|u_y|$, and $\tau = |\tau_{xy}|$, with 
\begin{equation}
    \tau_{xy} = (y-h)G + \mathcal{T}, \quad
    \mathcal{T} = \frac{1}{(1-h)^2}.
\end{equation}
Solving the quadratic \eqref{GNFconstquadratic1} for $\dot{\gamma}$, using the fact that $\mathrm{sgn}(u_y) = \mathrm{sgn}(\tau_{xy})$, then integrating, we find
\begin{multline}
    2Gu = \frac{\tau^2}{2} - (\delta+B)\tau - \frac{|GM|^2}{2} \\+ (\delta+B)|GM| + \frac{1}{2}(\tau+\delta-B)\sqrt{(\tau+\delta-B)^2 + 4\delta B} \\
    + 2\delta B\sinh^{-1}\left(\frac{\tau+\delta-B}{\sqrt{4\delta B}}\right)\\ - \frac{1}{2}(|GM|+\delta-B)\sqrt{(|GM|+\delta-B)^2+4\delta B} \\ 
    - 2\delta B\sinh^{-1}\left(\frac{|GM|+\delta - B}{\sqrt{4\delta B}}\right).
\end{multline}
Integrating $u$ across the fluid layer gives the flux, $q$, via
\begin{multline}
    2G|G|q = \frac{1}{6}\left[\mathcal{T}^3 - 2\tau_Z^3+|GM|^3\right] \\
    - \frac{1}{2}(\delta+B)\left[\mathcal{T}^2-2\tau_Z^2+|GM|^2\right] \\
    + G^2|M|h\left(\delta+B-\frac{1}{2}|GM|\right) \\
    + \frac{4(\delta B)^{3/2}}{3}\left[(\phi_\mathcal{T}^2+1)^{3/2}-2(\phi_Z^2+1)^{3/2}+(\phi_G^2+1)^{3/2}\right]\\
    +4(\delta B)^{3/2}\left[\phi_\mathcal{T}\sinh^{-1}(\phi_\mathcal{T}) - \sqrt{\phi_\mathcal{T}^2+1}\right]\\
    + 4(\delta B)^{3/2}\left[ - 2\phi_Z\sinh^{-1}(\phi_Z) + 2\sqrt{\phi_Z^2+1}\right]\\ 
    +4(\delta B)^{3/2}\left[\phi_G\sinh^{-1}(\phi_G) - \sqrt{\phi_G^2+1}\right]\\
    -2\delta Bh|G|\left[\sinh^{-1}(\phi_G) + \phi_G\sqrt{\phi_G^2+1}\right],
    \label{qGNF}
\end{multline}
where
\begin{equation*}
    M = h - \frac{\mathcal{T}}{G},\quad Z = \min\left[h,\max\left(0,M\right)\right],\quad
\end{equation*}
\begin{equation*}
    \tau_Z = |G||M-Z|,\quad 
    \phi_\mathcal{T} = \frac{\mathcal{T} - B + \delta}{(4\delta B)^{1/2}},\quad 
\end{equation*}
\begin{equation}
    \phi_Z = \frac{\tau_Z - B + \delta}{(4\delta B)^{1/2}},\quad 
    \phi_G = \frac{|GM| - B + \delta}{(4\delta B)^{1/2}}.\label{qGNFdefns}
\end{equation}


\section{Travelling waves in the large-$\mathcal{S}$ limit}\label{app:largeS}

In the limit $\mathcal{S}\gg1$, we can seek asymptotic solutions to the travelling-wave equation \eqref{traveleqn}. As in \S\ref{sec:travel}, we consider steady waves in periodic domains with length $L = 2\pi/k_m$, and neglect gravity, $\mG=0$. Given the dependence of $k_m$ on $\mathcal{S}$ \eqref{LSA_omr}, we introduce a scaled coordinate, $X = \mathcal{S}^{-1/2}\xi$, so that
\begin{equation}
    G = -\mathcal{S}^{3/2} \left(\frac{h_X}{(1-h)^3} + h_{XXX}\right) + O(1),\quad \mathcal{T}=O(1). \label{largeS:G}
\end{equation}
Similarly to other capillary flows such as collar or droplet translation under gravity \cite{kalliadasis1994drop,jensen_draining_2000,shemilt2025viscoplasticity}, we anticipate (partly motivated by numerical simulations) that the asymptotic solution is likely to be composed of a main wave body, where $h,X=O(1)$, a thin uniform film, $h\approx h_\infty\ll1$, ahead of and behind the wave body, and short intervening matching regions. Here, we do not pursue a full exposition of the matched-asymptotic
structure of the solution, but extract some insight from analysing the leading-order solution for $h$ in the main wave body.

\begin{figure}[t!]
    \centering
    \includegraphics[width=\linewidth]{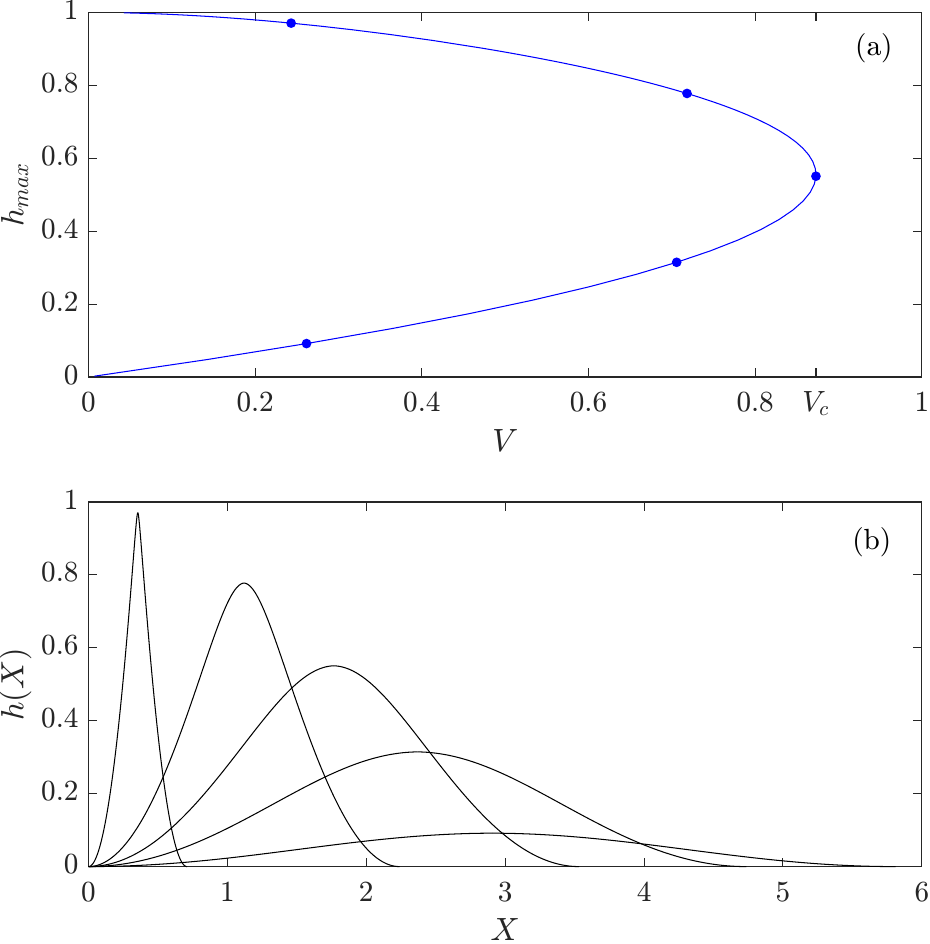}
    \caption{(a) Maximum height, $h_{max}$, against volume, $V$, of solutions to \eqref{largeS:heqn} for the leading-order wave shape, $h(X)$, in the limit $\mathcal{S}\gg1$. The maximum volume for which a solution exists is $V=V_c$. (b) Five example solutions, which correspond to the blue dots in (a).
    }
    \label{fig:largeS}
\end{figure}

We assume that $U\ll\mS^{3/2}$, which may be verified from numerical solutions, and we take the liquid flux to be equal to the Newtonian flux \eqref{qNewtonian}. For fixed $\mJ$, $B\rightarrow0$ as $\mS\rightarrow\infty$, so we expect viscoplastic layers with fixed $\mJ$ to behave like Newtonian layers in the limit $\mS\rightarrow\infty$, with the flux given by \eqref{qNewtonian} to leading order. At leading order in $\mS$, \eqref{traveleqn}, \eqref{qNewtonian} and \eqref{largeS:G} imply that $h$ must satisfy
\begin{equation}
    \frac{h_X}{(1-h)^3} + h_{XXX} =0,\label{largeS:heqn}
\end{equation}
reflecting a balance between the destabilising effect from air inertia and
stabilisation by surface tension. We solve this equation over the region
$[0,X_L]$, imposing $h = h_X = 0$ at the ends and the volume constraint
\begin{equation}
    \int_0^{X_L} h\,\mathrm{d}X = V,
\end{equation}
where $V$ is some prescribed constant and the 
width of the main body of the wave, $X_L$,
becomes determined as part of the solution. 
Figure \ref{fig:largeS} shows the resulting relation between the
peak height and volume, together with
several sample solutions. From figure \ref{fig:largeS}(a),
we observe that there is a maximum volume, $V_c\approx0.873$,
above which there are no solutions to \eqref{largeS:heqn}.
Given that most of the fluid layer becomes entrained into the
main wave body for $\mS\gg1$, we have $V\sim\bar{h}L=\bar{h}2\pi/k_m$.
Hence there is a maximum mean film thickness of $\bar{h}\approx0.119$
that can support steady travelling waves in this limit. In addition,
for $\bar{h}=0.1$, we found two disconnected branches of solutions to the full
travelling-wave equation \eqref{traveleqn} (see figure \ref{fig:Ntravel}a).
Indeed, figure \ref{fig:largeS}(a) implies that for $\bar{h}<0.119$,
there are two travelling-wave solutions with different peak
heights in the limit $\mS\to\infty$, which correspond to asymptotes
of the steady wave branches with finite $\mS$.

\section{Effect of gravity in the long-wave model}\label{app:grav}

\begin{figure}[t!]
    \centering
    \includegraphics[width=\linewidth]{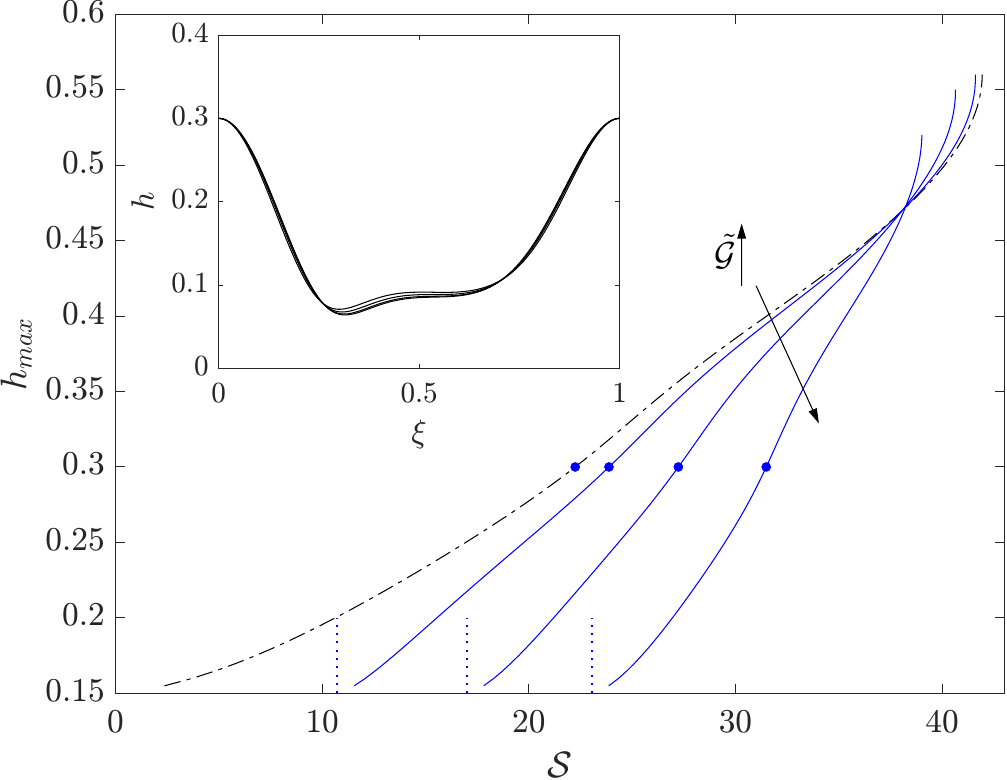}
    \caption{Maximum height of Newtonian travelling-wave solutions with $\bar{h}=0.15$ and various values of $\mG\in\{0,2000,8000,20000\}$. Dotted blue lines correspond to the minimum $\mS$ required for instability, for each value of $\mG$, according to \eqref{LSAkm}. 
      The inset shows sample solutions, corresponding to the
      blue dots in the main panel, all 
      with $h_{max}=0.3$. }
    \label{fig:travgrav}
\end{figure}

In this appendix, we briefly examine the effect of gravity on steady
travelling-wave solutions of the Newtonian
long-wave model in periodic domains with
$L=2\pi/k_m$. Figure \ref{fig:travgrav} shows solution branches on the
$(\mS,h_{max})-$plane for three values of $\mG$, all with
$\bar{h}=0.15$ (the gravity-less case is also shown).
As evident from \eqref{LSA_omr}, when $\mG>0$, there is no instability
when $\mS$ is too small, and the minimum value of $\mS$ required for
instability increases with $\mG$. Beyond that stability threshold,
the solution branches resemble that without gravity, and there
is a saddle node at a similar value of $\mS$ for each value of $\mG$
plotted in figure \ref{fig:travgrav}. Wave profiles possessing
the same peak height for different values of $\mG$ have similar shape
(see the inset in figure \ref{fig:travgrav}), despite the fact that the
values of $\mS$ corresponding to these solutions are quite different.
In fact, provided solutions do not proceed towards blow-up ($h\to1$),
the gravitational contribution to the pressure gradient
(the last term in \eqref{eq:GG}) is similar, but opposite,
to that provided by air inertia acting on a sloping free
surface (the first term on the right of \eqref{eq:GG}). Hence,
one expects that the effect of increasing $\mG$ mirrors that
of lowering $\mS$.
We conclude that the main qualitative impact of gravity
is to raise the minimum air speed for surface-wave instability,
or that required to observe some other feature of the nonlinear wave dynamics.

\bibliographystyle{elsarticle-num}
\bibliography{jfm.bib}

\end{document}